\documentclass[rmp,twocolumn,superscriptaddress,longbibliography,floatfix]{revtex4-2}

\usepackage[]{fontenc}

\usepackage{graphicx}
\usepackage{bm}
\usepackage{mathtools} 
\usepackage[caption=false]{subfig}
\usepackage{hyperref}
\usepackage{chngcntr} 
\usepackage{multirow}
\usepackage{subeqnarray}
\usepackage{wasysym}

\usepackage{amsmath}
\usepackage{amssymb}
\usepackage{framed}
\usepackage{soul}
\usepackage{color}
\usepackage{alltt}
\usepackage{pstricks,pst-node,pst-tree,pst-grad,pst-text,graphics}
\usepackage{dcolumn}
\usepackage{url}
\usepackage{cases}

\usepackage{ifthen}
\usepackage{afterpage}
\usepackage{comment}
\usepackage{supertabular}
\usepackage{xspace}
\usepackage{setspace}
\usepackage{textcomp}
\usepackage{tikz}
\usetikzlibrary{decorations.pathmorphing}
\usetikzlibrary{decorations.markings}
\usetikzlibrary{arrows,shapes,decorations,automata,backgrounds,petri}

\newcommand{\gpcomment}[1]{\PackageError{commentCommand}{Don't use comments in production LaTeX}{}}

\newcommand{\plaind}{\mathrm{d}}
\newcommand{\dint}[1]{\mathchoice{\!\plaind#1\,}{\!\plaind#1\,}{\!\plaind#1\,}{\!\plaind#1\,}}
\newcommand{\ddint}[1]{\ddintx{#1}{d}}

\newcommand{\ddintx}[2]{\mathchoice{\!\plaind^{#2}#1\,}{\!\plaind^{#2}#1\,}{\!\plaind^{#2}#1\,}{\!\plaind^{#2}#1\,}}

\DeclareFontFamily{U}{wncy}{}
\DeclareFontShape{U}{wncy}{m}{n}{<->wncyr10}{}
\DeclareSymbolFont{mcy}{U}{wncy}{m}{n}
\DeclareMathSymbol{\sha}{\mathord}{mcy}{"58}

\usepackage{dsfont}

\newcommand{\gpset}[1]{\mathds{#1}}

\newcommand{\canetset}[1]{{\mathchoice {\hbox{$\sf\textstyle #1\kern-0.4em #1$}}
{\hbox{$\sf\textstyle #1\kern-0.4em #1$}}
{\hbox{$\sf\scriptstyle #1\kern-0.3em #1$}}
{\hbox{$\sf\scriptscriptstyle #1\kern-0.2em #1$}}}}

\newcommand{\Rset}{\gpset{R}}

\def\nbZ{{\mathchoice {\hbox{$\sf\textstyle Z\kern-0.4em Z$}}
{\hbox{$\sf\textstyle Z\kern-0.4em Z$}}
{\hbox{$\sf\scriptstyle Z\kern-0.3em Z$}}
{\hbox{$\sf\scriptscriptstyle Z\kern-0.2em Z$}}}}

\newcommand{\gpvec}[1]{\mathbf{#1}}

\newcommand{\ivec}{\gpvec{i}}
\newcommand{\jvec}{\gpvec{j}}
\newcommand{\kvec}{\gpvec{k}}

\newcommand{\xvec}{\gpvec{x}}

\newcommand{\HC}{\mathcal{H}}

\newcommand{\OC}{\mathcal{O}}
\newcommand{\PC}{\mathcal{P}}

\newcommand{\half}{\mathchoice{\frac{1}{2}}{(1/2)}{\frac{1}{2}}{(1/2)}}

\newcommand{\Exp}[1]{\operatorname{exp}\left(#1\right)}
\renewcommand{\exp}[1]{\mathchoice%
{\mathrm{e}^{#1}}%
{\operatorname{exp}(#1)}
{\operatorname{exp}\left(#1\right)}%
{\operatorname{exp}\left(#1\right)}}

\newcommand{\elabel}[1]{\label{eq:#1}}
\newcommand{\eref}[1]{(\ref{eq:#1})}
\newcommand{\Eref}[1]{\mbox{Eq.~(\ref{eq:#1})}}
\newcommand{\Erefs}[1]{\mbox{Eqs.~(\ref{eq:#1})}}

\newcommand{\seclabel}[1]{\label{sec:#1}}

\newcommand{\Sref}[1]{Section~\ref{sec:#1}}

\newcommand{\flabel}[1]{\label{fig:#1}}

\newcommand{\Fref}[1]{Figure~\ref{fig:#1}}

\newcommand{\latin}[1]{{\it #1}}
\newcommand{\ie}{\latin{i.e.}\@\xspace}

\newcommand{\cf}{\latin{cf.}\@\xspace}

\newcommand{\etal}{\latin{et~al}.\@\xspace}
\newcommand{\vs}{\latin{vs}.\@\xspace}

\newlength \standardfigwidth
\DeclareMathAlphabet{\matheub}{U}{eur}{m}{n}

\usepackage{tikz}
\usetikzlibrary{decorations.pathmorphing}
\usetikzlibrary{decorations.markings}
\usetikzlibrary{arrows,shapes,decorations,automata,backgrounds,petri}
\usetikzlibrary{calc}
\usetikzlibrary{patterns}
\usetikzlibrary{decorations.text}

\newcommand{\ave}[2][]{\mathchoice%
{\left\langle #2 \right\rangle_{#1}}%
{\langle #2\rangle_{#1}}
{\langle #2\rangle_{#1}}%
{\langle #2\rangle_{#1}}}

\makeatletter
\newcommand{\creatX}[2][]{a^{\@ifempty{#1}{\dagger}{\dagger\,#1}}\@ifempty{#2}{}{(#2)}}
\makeatother
\newcommand{\creatDS}{\tilde{a}}
\makeatletter
\newcommand{\creatDSX}[2][]{\@ifempty{#1}{\creatDS}{\creatDS^{#1}}\@ifempty{#2}{}{(#2)}}
\makeatother

\makeatletter
\newcommand{\annihX}[2][]{a\@ifempty{#1}{}{^{#1}}\@ifempty{#2}{}{(#2)}}
\newcommand{\kernel}[1]{K\@ifempty{#1}{}{\,\!\!\left(#1\right)}}
\newcommand{\kernelDeri}[1]{K'\@ifempty{#1}{}{\left(#1\right)}}

\newcommand{\tableofappendixcontents}{\@starttoc{toa}}
\newcommand{\l@toact}[2]{\@dottedtocline{1}{1.5em}{5.5em}{#1}{#2}}
\newcommand{\l@toactspace}[2]{\ \hfil}
\makeatother

\newcommand{\APref}[1]{\mbox{App.~\ref{sec:#1}}}

\renewcommand{\exp}[1]{\mathchoice%
{e^{#1}}%
{\operatorname{exp}(#1)}%
{\operatorname{exp}(#1)}%
{\operatorname{exp}(#1)}}

\usepackage{xspace}

\renewcommand{\st}[1]{}

\makeatletter

\newcommand{\HMdensity}[2]{\rho_{#2}^{\@ifempty{#1}{}{(#1)}}}
\newcommand{\HMdensityBar}[2]{\overline{\rho}_{#2}^{\@ifempty{#1}{}{(#1)}}}
\newcommand{\HMdensityDot}[2]{\dot{\rho}_{#2}^{\@ifempty{#1}{}{(#1)}}}
\makeatother

\usepackage{textcomp}

\newcommand{\diffusion}{D}
\newcommand{\transDiffusion}{\diffusion_x}

\newcommand{\hot}{\text{h.o.t.}}

\newcommand{\invTemp}{\boldsymbol{\beta}_T}
\newcommand{\invTempCrit}{\boldsymbol{\beta}_{T,c}}
\newcommand{\MFT}{\text{\tiny MFT}}
\newcommand{\Ising}{\text{\tiny Ising}}
\newcommand{\eff}{\text{\tiny eff}}
\newcommand{\droplet}{\circ}
\newcommand{\corr}{\mathcal{C}}

\usepackage{color}
\definecolor{darkgreen}{rgb}{0,0.6,0}
\definecolor{darkblue}{rgb}{0,0,0.6}
\definecolor{darkred}{rgb}{0.6,0,0}
\definecolor{darkpurple}{rgb}{0.5,0,0.5}
\hypersetup{
bookmarksopen=true,
pdftitle="",
pdfauthor="", 
pdftoolbar=false, 
pdfstartview={FitH},		
pdfmenubar=true,			
pdfhighlight=/O,			
colorlinks=true,			
urlcolor=darkblue,
citecolor=darkblue,		
linkcolor=darkblue}

\makeatletter
\newcommand{\customlabel}[2]{%
   \protected@write \@auxout {}{\string \newlabel {#1}{{#2}{\thepage}{#2}{#1}{}} }%
   \hypertarget{#1}{\hspace{0pt}}
}
\makeatother

\begin{document}

\newcommand{\titleText}{Numerical investigation of the Brownian 
$q=2$ Potts Model}
\title{\titleText}

\author{Letian Chen}%
 \email{letian.chen16@imperial.ac.uk}
\affiliation{%
Department of Mathematics
and Centre of Complexity Science, 
Imperial College London, London SW7 2AZ, United Kingdom}%

\author{Gunnar Pruessner}
\email{g.pruessner@imperial.ac.uk}
\affiliation{%
Department of Mathematics
and Centre of Complexity Science, 
Imperial College London, London SW7 2AZ, United Kingdom}%

\date{April 2024}

\begin{abstract}
In active matter, such as the Vicsek Model of flocking, particles possesses an internal degree of freedom, such as their director, which is subject to interaction with other particles, provided they are within a certain range. In an effort to understand better the interplay between spatial and internal degrees of freedom, we study numerically a variation of the $q=2$ Potts Model on and off the lattice, where particles are additionally subject to Brownian motion. The lack of a feedback of the internal degrees of freedom to the spatial degrees of freedom renders this model generically non-equilibrium. We confirm previous work that showed that the static exponents of the phase transition are unaffected by the diffusion. In contrast to previous work, we show that the formation of ordered clusters is not undermined by diffusion, but should rather be thought of as an effective form of interaction. We demonstrate how our numerical findings can be understood on the basis of the well-established Model~A, B and C: Off lattice, the Brownian $q=2$ Potts Model is Model~C. On the lattice, it is Model~A with an additional (irrelevant, conserved) Model~B noise.
\end{abstract}

                              
\maketitle

\newcommand{\longtodo}[1]{\todo[inline,size=\tiny]{#1}}

\section{Introduction}
\seclabel{introduction}

One major controversy surrounding the Vicsek Model \cite{viscek_1995} and other models \cite{Solon_2013,shaebani_computational_2020,Deseigne_2010,Romanczuk_2009,Holger_2022,Das_2024,Caprini_2023} of flocking remains the questions whether there is an ordered state in two dimensions, the nature of that ordered state and the characteristics of a possible phase transition into that state. What makes the question so relevant and interesting is that no such ordered state can exist in equilibrium either on nor off the lattice, because of the Mermin-Wagner theorem \cite{MerminWagner_1966} and extensions to it \cite{Tasaki_2020}, which imply that \emph{in equilibrium} no long-range order exists in two dimensions between continuous degrees of freedom subject to short-ranged interaction. The order observed consistently in the Vicsek Model \cite{viscek_1995} appears to be an oddity, even when allowed by its non-equilibrium nature which protects it from the spell of the Mermin-Wagner theorem.

On the other hand, models with discrete symmetries are known to display long-range order in two dimensions. The Ising model and equally the 
$q=2$ Potts Model are particularly well understood in this respect \cite{Wu_1982}. 
In the following, we study the $q=2$ Potts Model with diffusion, the \emph{Brownian $q=2$ Potts Model} \cite{Woo_2022}, as a minimal variation of the conventional $q=2$ Potts Model that brings it closer to the Vicsek Model, but without some of the major complications in the Vicsek Model: The diffusion takes the Potts Model out of equilibrium, but we do not implement any self-propulsion and alignment interaction has no effect on the spatial degree of freedom. 
As \citeauthor{Tasaki_2020} (\citeyear{Tasaki_2020}) points out, the non-equilibrium nature might bring about order in the Vicsek and related models, which suggests that anything violating equilibrium might be a relevant perturbation.
In the following, we add diffusion to a classical model that displays a well-understood order-disorder transition in the absence of diffusion.
\emph{We will focus on the question to what extent the diffusive motion of the interaction partners poses a relevant perturbation.}

The local order parameter of the $q=2$ Potts Model is scalar and does not obey a continuous symmetry so that in the limit of vanishing diffusion, we expect to recover the standard Ising Model, albeit with disorder when the interaction partners are uniformly distributed. Further, as particles are not self-propelled, but purely diffusive, there is no feedback of the internal state, \ie the spin orientation, to the movement of the particle. As a result 
the stochastic equations of motion are not expected to carry an advection term as, say, the Toner-Tu equations do \cite{TonerTu:1995}.
there is no advective term present. We ask: How is the $q=2$ Potts Model modified by the presence of such a non-equilibrium diffusive term?

The basic features of the Brownian $q=2$ Potts Models studied in the literature and further detailed below are: 1) Particles have an ``internal'' Potts degree of freedom subject to Glauber \cite{Woo_2022,Loos_2023,Bandyopadhyay_2024,Solon_2013,Solon_2015} or Metropolis \cite{Woo_2024}
updating as a form of spin-interaction with a local neighbourhood. 2) Particles are subject to diffusion off-lattice or, in the present work nearest-neighbour swapping on the lattice, irrespective of the internal Potts state. Because the swapping and diffusion occurs without reference to the internal Potts state, the models are generically out of equilibrium. Forces are not reciprocal: Particles move freely, even when their interaction makes them align with their neighbourhood. This lack of mechanical consistency would allow for infinite energy extraction whereby particles are subject to an alignment force when they are close, can be separated without performing work, then flipped without work and re-align once brought close together again. 

The non-equilibrium nature of the setup can also be observed in the following \emph{Gedankenexperiment}: If there is no mechanical feedback of the spin orientation to the movement of particles, their trajectories approaching each other are indistinguishable from their trajectories departing from each other. However, as they get close, they will tend to align. As a result, their will be plenty of forward trajectories of the form: approach misaligned, align, depart aligned. As particles align as they get close, they typically depart aligned, so that the reverse trajectory is clearly less likely, breaking time-reversal symmetry. In equilibrium, the same construction cannot be made, because particles attract each other as they are aligned and separate (more easily) once misaligned, boosting the frequency with which misaligned particles are seen departing each other.

Solon and Tailleur \cite{Solon_2013,Solon_2015} have studied a Brownian Ising Model on the lattice. 
As this model was inspired by the Vicsek Model \cite{viscek_1995}, particles, equipped with a spin, self-propel according to their spin orientation. Without any self-propulsion particles are free to diffuse. Spins are subject to an all-to-all interaction when they reside on the same site. Otherwise, they do not interact. Nevertheless, this model displays long-ranged ordering as the diffusion establishes effectively interaction from site to site. Although observed only in the particle density as the control parameter, Solon \etal 's setup displays in two dimensions the standard Ising phase transition. At first sight, this comes as a surprise, because the local interaction is all-to-all and thus inspired by mean-field models. What is more striking is the r{\^o}le of the diffusion. On the one hand, it mediates the interaction between nearest neighbouring sites, as particles with aligned spins ``invade'' neighbouring sites and ``win them over''. On the other hand, the diffusion also results in a loss of local magnetisation as particles leave their aligned neighbourhoods.

This mechanism is key in the setup studied by Woo, Rieger and Noh \cite{Woo_2022} who studied a $q=2, \dots 8$ Potts Model off the lattice with particles subject to random motion that amounts essentially to diffusion. Again, there is no spatial force acting on the particles as their motion is unaffected by the state of mutual alignment, but there is an alignment force acting depending on their proximity, which is being accounted for by heat-bath (Glauber) dynamics. They find that for $q=2$, essentially the Ising Model, the \emph{static} characteristics of the phase transition are unchanged by the diffusion, although there is clear evidence that the \emph{dynamical} exponent is reduced from $z=2.1665(12)$ of Model~A \cite{Nightingale_1996} to $z=2$. The effect for $q=3$ and $q=4$ appears to be more striking, as the static exponents deviate more clearly from exactly known values on the lattice. Finally, $q=5, \dots 8$ display a continuous phase transition in the off-lattice Brownian model by Woo \etal compared to the first order transitions established on the lattice \cite{Baxter_1973}.

What is so striking about Woo \etal's result \cite{Woo_2022} for the $q=2$ Potts Model is that, apparently, the diffusion of the interaction partners is relevant (enough) to change the dynamical exponent $z$ from $2.1665(12)$ to that of plain diffusion, $z=2$, suggesting that the dynamics is diffusion dominated. On the other hand, the static critical behaviour is unaffected  by the diffusion, so that standard Ising Model, \ie the static exponents of Model~A are observed. Surprisingly, to explain the phenomenology of the Brownian Potts Model, \citeauthor{Woo_2022} develop a reasoning that suggests that the Brownian $q=2$ Potts Model should display mean-field behaviour \cite[][Sec. IV.]{Woo_2022}, \Sref{droplet_picture}.
While Woo \etal do not observe such mean-field behaviour in their numerics, we wonder whether this is due to the particular parameter and size regime studied.

In the following, we focus our attention at the $q=2$ Potts Model, on and off the lattice, with the aim to either find the mean-field behaviour on sufficiently large scales or
identify the mechanism by which only the dynamical exponent $z$ but not the static exponents are modified compared to Model~A \cite{HohenbergHalperin_1977} in the presence of diffusion.
We will use the standard models and their classification by \citeauthor{HohenbergHalperin_1977} throughout: Model~A is the simplest dynamical version of $\varphi^4$-theory, in the scalar form with Glauber or, equivalently, heat-bath dynamics, known to capture both $q=2$ Potts and Ising dynamical critical behaviour \cite{Tauber_2008}.
Model~B plays the same r{\^o}le but with a conserved order parameter. Model~C may be thought of as the simplest extension of Model~A with the order parameter coupling to a conserved quantity.
After introducing the models in \Sref{model_defs} and the effective field theories they might be cast in, namely Model~C, and Model~A with Model~B noise, we report the results we found numerically in \Sref{results}, providing further analytical context for our findings and conclusions in \Sref{discussion}.

\begin{figure*}[!t]
\centering
\subfloat[\flabel{DIM_snapshot} Off-lattice model: Linear size $L=50$, particle density $\rho=2$, interaction range $r_0=1$ and hopping length $v_0=2$. The radius of the particles corresponds to the interaction range. ]{%
  \includegraphics[width=0.9\textwidth]{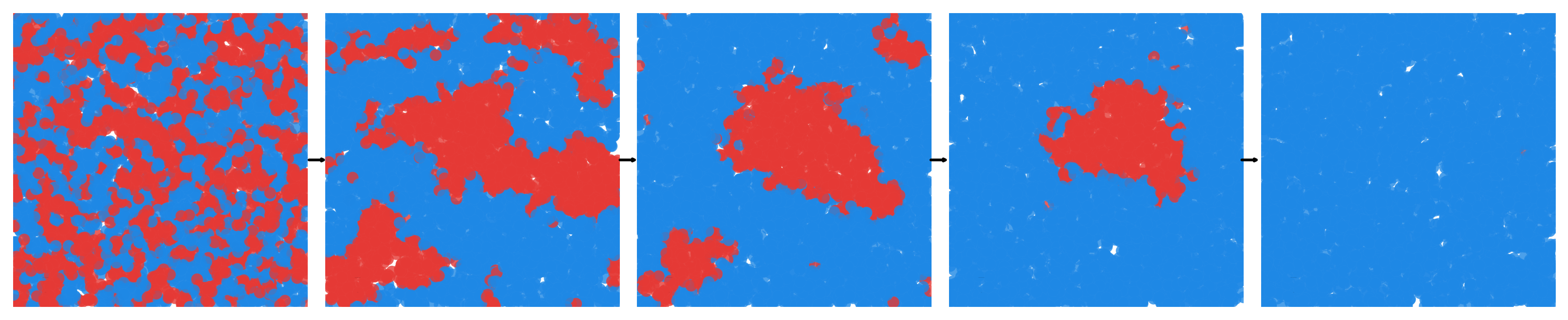}%
}
\\[1em] 
\subfloat[\flabel{SIM_snapshot} On-lattice model: Linear size $L=50$.]{%
  \includegraphics[width=0.9\textwidth]{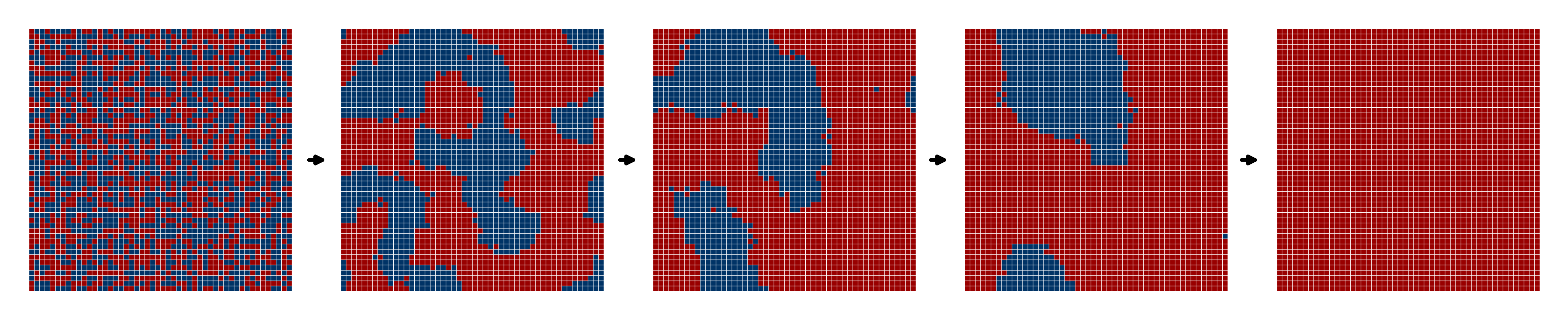}%
}%
\caption{\flabel{snapshot} Time-evolution snapshots of the off-lattice model, \Sref{off_lattice_particle}, and the on-lattice model, \Sref{on_lattice_model},  after a quench from $T=\infty$ into the ordered phase. Blue indicates spin-up, whereas red indicates spin-down. }
\end{figure*}

\section{Model definitions}\seclabel{model_defs}
In the present work we are studying 
two models, namely the \emph{off-lattice Brownian $q=2$ Potts Model}, 
to our understanding almost identical to that 
studied by Woo, Rieger and Noh \cite{Woo_2022} and the \emph{on-lattice $q=2$ Potts Model}, with the specific aim to 
characterise their critical behaviour and 
determine the universality classes their respective transitions belong to.
We study these models numerically and develop, alongside, a field-theoretic perspective on their particular features.

\subsection{Off-lattice particle model}\seclabel{off_lattice_particle}
The off-lattice model is defined as follows. In a two-dimensional box of linear extent $L$ and periodic boundary conditions, $N = \rho L^2$ particles of density $\rho$ indexed by $i\in\{1,\ldots,N\}$ have positions $\xvec_i\in[0,L)^2$. Initially, those positions may be chosen to be taken from a uniform distribution. As there is no mechanical feedback on the position, not even hard-core repulsion, the stationary state is bound to be uniform, just like that of an ideal gas \cite{Kardar_2007}. Each particle further has an internal degree of freedom in the form of 
a state variable $\sigma_i$ that can take any of $q$ values. For $q=2$, which we are focusing on in the following, it is instructive to think of this degree of freedom as 
an Ising spin $\sigma_i\in\{-1,1\}$. As argued below, the treatment of the Ising spins as just that or as $q=2$ Potts degrees of freedom, introduces a subtle difference, which turns out to amount to a shift of temperature.

When the spins are within distance $r_0$, their Potts state interact according to a dynamics characterised by the ``Potts-Hamiltonian''
\begin{equation}\elabel{def_Hamiltonian}
\HC(\{\sigma_i\}) = -J \sum_{i=1}^{N-1}\sum_{j=i+1}^N \theta\left(
1-\frac{|\xvec_i-\xvec_j|}{r_0}\right) 
\delta_{\sigma_j,\sigma_i}
\end{equation}
where $\{\sigma_i\}$ denotes the specific, given set of the states of all spins in the entire system, $J$ is an interaction strength in the following assumed to be unity, $\theta(z)$ denotes the Heaviside $\theta$-function and $\delta_{\sigma_j,\sigma_i}$ the Kronecker $\delta$-function. 
The distance $|\xvec_i-\xvec_j|$ is taken with periodic boundary conditions, so that $|\xvec_i-\xvec_j|$ is in fact the shortest path between particle $i$ and $j$ on the torus.
The summation limits are chosen as to sum over every pair of particles once, resulting in $N(N-1)/2$ terms. The Hamiltonian \Eref{def_Hamiltonian} implements a Potts-Model \cite{Baxter_1973} rather than the standard Ising Model that would see $\delta_{\sigma_j,\sigma_i}$ replaced by $\sigma_j\sigma_i$. Of course, as far as the critical behaviour is concerned, this is not expected to make any material difference. However, rewriting 
\begin{equation}\elabel{Potts_vs_Ising}
    \delta_{\sigma_j,\sigma_i}
= \half \sigma_j\sigma_i + \half
\end{equation}
points at a subtle difference between Ising and Potts. Firstly, the Potts Hamiltonian \Eref{def_Hamiltonian} carries, effectively, a factor $1/2$ that an Ising Hamiltonian would not be carrying, so that the Ising Model at temperature $2T$ is equivalent to the Potts Model at temperature $T$. Secondly, there is a constant shift by $1/2$, which means that the Hamiltonian \Eref{def_Hamiltonian} is shifted with respect to an Ising Hamiltonian by an amount proportional to the number of interaction partners. When this number varies, \ie it is configuration-dependent, it cannot be absorbed by the normalisation of the probability, as would happen in a Boltzmann factor. However, in the present case of the dynamics, to be discussed in detail below, not respecting mechanical consistency, the dynamics does not account for this shift, so that \emph{the present Brownian $q=2$ state Potts Model is equivalent to the Brownian Ising Model}. The shift would indeed enter in an equilibrium version of the present model, lowering the energy in the Potts model by $Jq/2$ and therefore favouring configurations with many interaction partners, similar to the vision cone model by \citeauthor{Loos_2023} (\citeyear{Loos_2023}), is non-equilibrium for being non-reciprocal.

The Hamiltonian enters in a heat-bath algorithm, whereby spins are updated random sequentially as follows: Choosing particle $i$ we determine 
\begin{equation}\elabel{def_hi_off_lattice}
h_i^\pm = \sum_{j=1}^N \theta\left(1-
|\xvec_i-\xvec_j|/r_0\right) \delta_{\sigma_j,\pm1}
\end{equation}
akin to the local (molecular) field spin $i$ is subjected to, so that the Hamiltonian with $\sigma_i=\pm1$ is $\HC=\HC_0-Jh_i^\pm$, with $\HC_0$ containing all terms in $\HC$ that do not contain spin $\sigma_i$. 
With $\invTemp$ the inverse temperature,
spin $i$ is then updated by choosing $\sigma_i=+1$ with probability $\exp{\invTemp J h_i^+}/(\exp{\invTemp J h_i^+}+\exp{\invTemp J h_i^-})$ and $\sigma_i=-1$ otherwise. Glauber dynamics amounts to the same outcome, although it is formulated in terms of the probability to \emph{flip} a spin.

If the particles' positions $\xvec_i$ were fixed, the $\theta\left(1-|\xvec_i-\xvec_j|/r_0\right)$ could be cast in a fixed interaction matrix and the update scheme described above is the well-known heat-bath dynamics \cite{Krauth_2006} that locally thermalises every spin, producing equilibrium probability $\propto\exp{-J\invTemp\HC}$.

After thermalising all spins sequentially, spin positions are updated by incrementing every $\xvec_i$ by distance $v_0$ in a uniformly and randomly and independently chosen direction, say $\vartheta_i$ chosen uniformly from $[0,2\pi)$ and $\xvec_i$ incremented by $(v_0 \cos(\vartheta_i),v_0\sin(\vartheta_i))$, applying periodic boundary conditions such that $\xvec_i\in[0,L)^2$ is maintained. New $\vartheta_i$ are chosen in each such update. Time $t$ advances by one unit after the thermalisation of all $\sigma_i$ and the update of all $\xvec_i$.

Because $\xvec_i$ sample a uniform spatial distribution in $[0,L)^2$, random sequential or deterministically sequential updates, whereby particles are visited in a fixed sequence, are not expected to make a noticeable difference. 
Noh \etal \cite{Woo_2022} seem to has used a parallel updating scheme. 

The dynamics described above is generically far-from-equilibrium, because of the non-reciprocal way spatial and spin degrees-of-freedom are coupled: If they are close enough, spins effectively ``feel a force'' as far as their spin-variable is concerned, but they do not ``feel a force'' as far as their displacement is concerned, no matter whether their spins are aligned or not.  \Fref{DIM_snapshot} illustrates the dynamical evolution of the quenching process in the off-lattice model.

\subsubsection{Observables}\seclabel{observables}
To characterise any possible phase transition, we introduce the instantaneous order parameter
\begin{equation}\elabel{def_m}
    m = \frac{1}{N} \sum_{i=1}^N \sigma_i
\end{equation}
whose expectation, $\ave{m}$, however, will vanish by symmetry in any finite system. As routinely used in the literature \cite{LandauBinder_2005}, numerically we therefore resort to estimating $\ave{|m|}$, which scales like $(\invTemp-\invTempCrit)^{\beta}$ in the ordered phase, $\invTemp>\invTempCrit$, where $\invTempCrit$ is the critical inverse temperature. The standard Ising exponent is $\beta_\Ising=1/8$ and the standard mean-field exponent is $\beta_\MFT=1/2$. At $\invTemp=\invTempCrit$ the order parameter displays finite size scaling $\ave{|m|}\propto L^{-\beta/\nu}$ with exponent $\nu_\Ising=1$ and $\nu_\MFT=1/2$.

Similarly, the susceptibility 
\begin{equation}
\elabel{ture_susceptibility}
    \chi=N (\ave{m^2}-\ave{|m|}^2)
\end{equation}
scales like $(|\invTemp-\invTempCrit|)^{-\gamma}$ and displays finite size scaling $\propto L^{\gamma/\nu}$ with $\gamma_\Ising=7/4$ and $\gamma_\MFT=1$. For all numerical simulations conducted in this paper, the effective $\invTempCrit(L)$ for each system size $L$ is defined as the value of $\invTemp$ at which the susceptibility $\chi$ reaches its maximum.

As for time-dependent observables, we use a variant of the two-time correlation function, $\ave{m(t_0+t) m(t_0)}$
which in the stationary state is independent of $t_0$. The product of the expectations $\ave{m(t_0+t)}$ and  $\ave{m(t)}$ that are normally subtracted from $\ave{m(t_0+t) m(t_0)}$ to render it a connected correlation function, can be omitted in finite systems as both vanish. However, numerically better behaved \cite{Anderson:1971} is often the symmetric estimator
\begin{multline}
\elabel{def_twoTimeCorr}
    \corr(t) = 
    \frac{1}{M-t} \sum_{t'=1}^{M-t} m_{t'+t} m_{t'} \\ -
    \left( \frac{1}{M-t} \sum_{t'=1}^{M-t} m_{t'+t} \right)
    \left( \frac{1}{M-t} \sum_{t'=1}^{M-t} m_{t'} \right)
\end{multline}
where $m_{t'}$ refer to the average \Eref{def_m} taken at time $t'=1,2,\ldots,M$ in a total set of $M$ samples.
Calculating the two-time correlation function for all $t=1,2,\ldots,M$ directly from \Eref{def_twoTimeCorr} requires time of order $M^2$, which  can be computationally very expensive. To accelerate this process, we first compute the Fourier transform $\tilde{\corr}(\omega)$ and then obtain $\corr(t)$ by inverting the Fourier transform. By using the fast Fourier transform (FFT) algorithm, the computation time is reduced to order $M \log M$ \cite{newman_1999}.
To leading order, we expect $\corr(t)$ to decay like $\corr(t)=\corr(0)\exp{-t/\tau}$, so that $\tau$ can be estimated from the integral of $\corr(t)$ over those $t$ before $\corr(t)$ is negative for the first time, considered to be the time when the estimate $\corr(t)$ becomes unreliable. 
We can then extract the dynamical critical exponent $z$ from the correlation time $\tau$ using finite size scaling, $\tau\propto L^z$.

\subsubsection{The droplet picture}\seclabel{droplet_picture}
As discussed by \citeauthor{Woo_2022} the diffusion of the particles seems to spoil the ordering and formation of clusters of particles with the same spin orientation. The argument goes as follows. In equilibrium, the time scale of the ordering of a cluster of size $\ell$ is $\propto\ell^z$, characterised by the dynamical exponent $z$ which in $\varphi^4$-theories is generally bounded from below by $2$. In equilibrium, Glauber dynamics produces the Model~A exponent $z\approx 2.1665(12) $ \cite{Hinrichsen_1998,Nightingale_1996}. Assuming for a moment that a sufficiently dense system with sufficiently small diffusion constant $\transDiffusion$ displays essentially equilibrium behaviour, clusters of size $\ell$ will thus require time $A\ell^z$ to form, with some amplitude $A$. However, as their constituents carry on diffusing, one might expect such clusters to disintegrate over time $\ell^2/\transDiffusion$, thereby limiting the maximum time for a cluster to form by $\ell^2/\transDiffusion$, so that the maximum cluster size is $\ell_\droplet=(A\transDiffusion)^{1/(2-z)}$. As $z>2$, any finite diffusion constant $\transDiffusion$ thus limits the formation of arbirtarily large clusters, and one might expect the transition to be that of a perfectly mixed system. The exponents for that are calculated in \APref{randomNeighbourIsing} and, as expected, turn out to be those of standard mean-field theory, in particular
$\ave{|m|}\propto(\invTemp-\invTempCrit)^{\beta_\MFT}$
and
$\chi \propto |\invTemp-\invTempCrit|^{-\gamma_\MFT}$.

As far as finite size scaling is concerned, the random neighbour model in \APref{randomNeighbourIsing} displays the behaviour of the Ising Model at the upper critical dimension $d_c=4$, but without the logarithmic corrections normally encountered there \cite{LeBellac_1991}. For finite size scaling, the link between the random neighbour model and mean-field theory is somewhat subtle: Standard finite size scaling is in the linear extent $L$, whereas the random neighbour model shows scaling in the number of sites $N$, as it does not have notion of space and linear extent. For example, the random neighbour model predicts
$\ave{|m|}\propto N^{-1/4}$ (with $4$ coming from the next even order in the effective potential of the magnetisation density when the quadratic vanishes). When a well-mixed system in $d$ dimensions shows this behaviour at constant particle density, this translates to 
$\ave{|m|}\propto L^{-d/4}$. On the other hand, at $d=4$ the standard Ising Model displays mean-field scaling $\ave{|m|}\propto L^{-1}\propto N^{-1/4}$ with additional logarithmic corrections ignored.
This scaling coincides with the random neighbour model only as far as the scaling in $N$ is concerned, but, unless $d>4$, not as far as the scaling in $L$ is concerned. To make the connection, one has to express the standard mean-field scaling in terms of $N\propto L^4$ and then translate that to the non-standard scaling in $L\propto N^{1/d}$, on account of the mean-field behaviour being observed in a dimension where non-mean-field behaviour is normally expected.

What at first appears to be a triviality, is not so trivial at closer inspection, because a mean-field theory, valid when all spins interact with all other spins, the theory at $d=4$, where fluctuations and correlations become essentially irrelevant, and the random neighbour model of \APref{randomNeighbourIsing} are simply not the same. 
To summarise, in $d=2$, the random neighbour model predicts
close to the critical point
$\ave{|m|}\propto (\invTemp-\invTempCrit)^{1/2}$
and
$\chi \propto |\invTemp-\invTempCrit|^{-1}$, and for finite size scaling
$\ave{|m|}\propto L^{-1/2}$
and
$\chi \propto L$.

The argument above about the destructive effect of the diffusion on the ordering hinges crucially on the diffusive dynamics countering the ordering. However, this is not necessarily the case, as can be seen in the active Ising Model as introduced by \citet{Solon_2013,Solon_2015}, where the \emph{sole} form of nearest neighbour interaction is due to diffusion. While therefore the diffusion clearly blurs domainwalls, one might rather think of it as an effective extension of the interaction range.

\subsection{Off-lattice field theory}\seclabel{ModelC}

In \emph{Model~C} \cite{HohenbergHalperin_1977}, the local order parameter $\phi(\xvec,t)\in\Rset$ couples to another (density) field $\rho(\xvec,t)$, to be related to the way the spins above couple to other spins in their neighbourhood. For the stationary distribution to be Boltzmann and thus known, the stochastic equations of motion of \emph{both} fields, $\phi$ and $\rho$, need to be governed by the same thermal noises and the same Hamiltonian.

Specifically, the stochastic equations of motion of Model~C are 
\begin{subequations}
\elabel{eom_ModelC}
\begin{align}
\dot{\phi}(\xvec,t) & = 
-\Big(
r \phi(\xvec,t) + \frac{u}{3!} \phi^3(\xvec,t)
\Big) 
+ \nu \nabla^2 \phi(\xvec,t) \nonumber \\
\elabel{eom_ModelC_phi}
& \qquad- g \rho(\xvec,t) \phi(\xvec,t)
+ \xi(\xvec,t)\\
\elabel{eom_ModelC_rho}
\dot{\rho}(\xvec,t) & = \nabla^2 \Big( \transDiffusion  \rho(\xvec,t) + w \frac{g}{2}\phi^2(\xvec,t) \Big) + \zeta(\xvec,t)
\end{align}
\end{subequations}
where $\phi(\xvec,t)$ is the local magnetisation (density), couplings $r$ and $u$ parameterise the local potential in the well-known Landau-Ginzburg-Wilson form, $\nu$ is the interaction strength in direct space and $g$ describes the coupling to another field, $\rho(\xvec,t)$, whose dynamics keeps it conserved. In the dynamics of $\phi(\xvec,t)$, the field $\rho(\xvec,t)$ has the effect of a perturbation of the mass $r$.

The field $\rho(\xvec,t)$ is often thought of as an energy \cite[][p 113, 208]{Tauber_2008}, but may as well be a density of mobile impurities \cite[][p 450]{HohenbergHalperin_1977}. In the present work, we consider it as the particle density \emph{fluctuation} about a fixed background, so that $\rho$ does not need to have the same sign throughout the system. Its evolution is governed by the translational diffusion $\transDiffusion$ and, crucially, by a coupling to the squared magnetisation density $\phi^2$. This term is part of the definition of Model~C, but no such mechanical feedback is present in the Brownian $q=2$ Potts Model of \Sref{off_lattice_particle}.

The two thermal noises $\xi(\xvec,t)$ and $\zeta(\xvec,t)$ have correlators 
\begin{subequations}
\elabel{ModelC_noise_correlators}
\begin{align}
\elabel{ModelC_noise_correlators_xi}
\ave{\xi(\xvec,t)\xi(\xvec',t')} &= \Gamma^2 \delta(\xvec-\xvec')\delta(t-t')\\
\ave{\zeta(\xvec,t)\zeta(\xvec',t')} &= - \gamma^2 \nabla^2 \delta(\xvec-\xvec')\delta(t-t')
\ .
\end{align}
\end{subequations}
The equations of motion \Erefs{eom_ModelC} can be derived from the Hamiltonian
\begin{multline}\elabel{Hamiltonian_ModelC}
    \HC_C = \int \ddint{x} \frac{r}{2}\phi_0^2(\xvec) + \frac{u}{4!} \phi_0^4(\xvec) + \frac{\nu}{2}(\nabla \phi_0(\xvec))^2 \\
    + \frac{g}{2}\phi_0^2(\xvec)\rho_0(\xvec) + \frac{1}{w}\frac{\transDiffusion}{2} \rho_0^2(\xvec)
\end{multline}
demanding the dynamics 
\begin{subequations}
\elabel{ModelC_Langevin_from_Hamiltonian}
\begin{align}
\elabel{ModelC_Langevin_from_Hamiltonian_phi}
    \dot{\phi}(\xvec,t) &= -
    \left.\frac{\delta\HC_C }{\delta \phi_0(\xvec)}\right|_{\phi_0(\xvec)\equiv\phi(\xvec,t)}  + \xi(\xvec,t)\\
    \dot{\rho}(\xvec,t) &= w \nabla^2 
    \left.\frac{\delta\HC_C}{\delta \rho_0(\xvec)}\right|_{\rho_0(\xvec)\equiv\rho(\xvec,t)}  + \zeta(\xvec,t)
\end{align}
\end{subequations}
and all degrees of freedom will have the Boltzmann equilibrium stationary state $\propto\exp{-\HC_C/(\Gamma^2/2)}$, 
provided the noises are ``the same", in the sense that
\begin{equation}
    \ave{\zeta(\xvec,t)\zeta(\xvec',t')} = - w \nabla^2 \ave{\xi(\xvec,t)\xi(\xvec',t')} \ .
\end{equation}
By comparison to \Erefs{ModelC_noise_correlators} this is the case provided only that $w=\gamma^2/\Gamma^2$, which is the choice to be made to render \Erefs{eom_ModelC} the original Model~C \cite{HohenbergHalperin_1977}.
To put it more succinctly: \Erefs{eom_ModelC} and \eref{ModelC_noise_correlators} display \emph{equilibrium} fluctuations and, thus, equilibrium static features, provided $w=\gamma^2/\Gamma^2$ and therefore setting $w=0$ in \Eref{eom_ModelC_rho} in order to remove the undesired coupling of the local magnetisation density to the density field does not appear to be an option to bring Model~C closer to our off-lattice model, unless fluctuations of the density field always vanish suitably, as $\gamma^2=w\Gamma^2$.

That \Eref{Hamiltonian_ModelC} produces standard \emph{static} $\varphi^4$ theory can be seen by integrating out the density field $\rho$, which, after completing squares in the Hamiltonian \Eref{Hamiltonian_ModelC}, 
\begin{equation}
    \frac{g}{2}\phi_0^2\rho_0 + \frac{1}{w}\frac{\transDiffusion}{2} \rho_0^2 =
    \frac{1}{w}\frac{\transDiffusion}{2} \left(
    \rho_0^2 + \frac{2w}{\transDiffusion}\frac{g}{4} \phi_0^2 
    \right)^2 - 
    \frac{2w}{\transDiffusion}\left(\frac{g}{4}\right)^2\phi_0^4
\end{equation}
shifts $u/4!$ to $u/4!-2w(g/4)^2/\transDiffusion$ in an effective Landau-Ginzburg-Wilson Hamiltonian \cite{Tauber_2008,LeBellac_1991}. The dynamical critical exponent of the order parameter, $z$, however, is affected by the presence of the additional field $\rho$, resulting in the exact relationship $z=2+\alpha/\nu$ \cite[][p 115, 216]{Tauber_2008} in the case of a scalar field $\phi$, \ie $z=2$ in $d=2$, as $\alpha$ vanishes exactly.

Although the Langevin \Erefs{eom_ModelC} can be written in the form \Erefs{ModelC_Langevin_from_Hamiltonian} only when $w=\gamma^2/\Gamma^2$, \citeauthor{Akkineni_2004} found that the critical exponents characterising the transition into the ordered state do not depend on $w$, suggesting that ``detailed balance becomes effectively restored at the phase transition''.
We believe that the off-lattice particle model of \Sref{off_lattice_particle} can be thought of as Model~C, with standard Ising static exponents and $z=2$. The subtleties of this mapping are relegated to the discussion in \Sref{diagrammatics_ModelC}.

The effective diffusion of particles introduces a form of disorder, which in the limit of $v_0\to0$ is effectively quenched. To separate the effect of a varying number of neighbours from that of the randomisation of the interaction partners, we further introduce the following lattice model.

\begin{figure*}[!t]
\centering  
\subfloat[\flabel{FSS_mag} Finite size scaling of the magnetization $\langle|m|\rangle$.]
{\includegraphics[width=1\columnwidth]{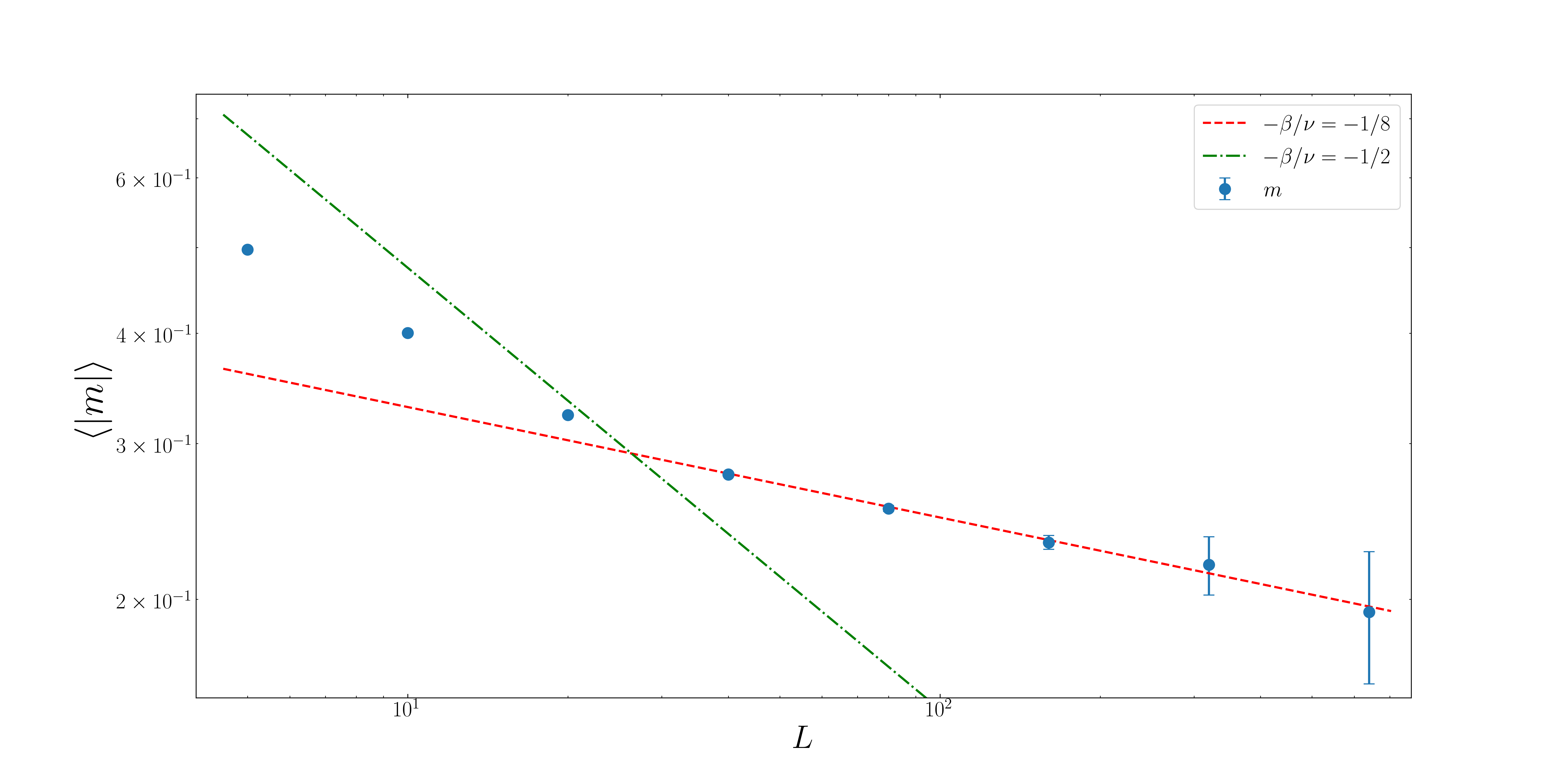}}
\hfill
\subfloat[\flabel{FSS_chi} Finite size scaling of the susceptibility $\chi$.]
{\includegraphics[width=1\columnwidth]{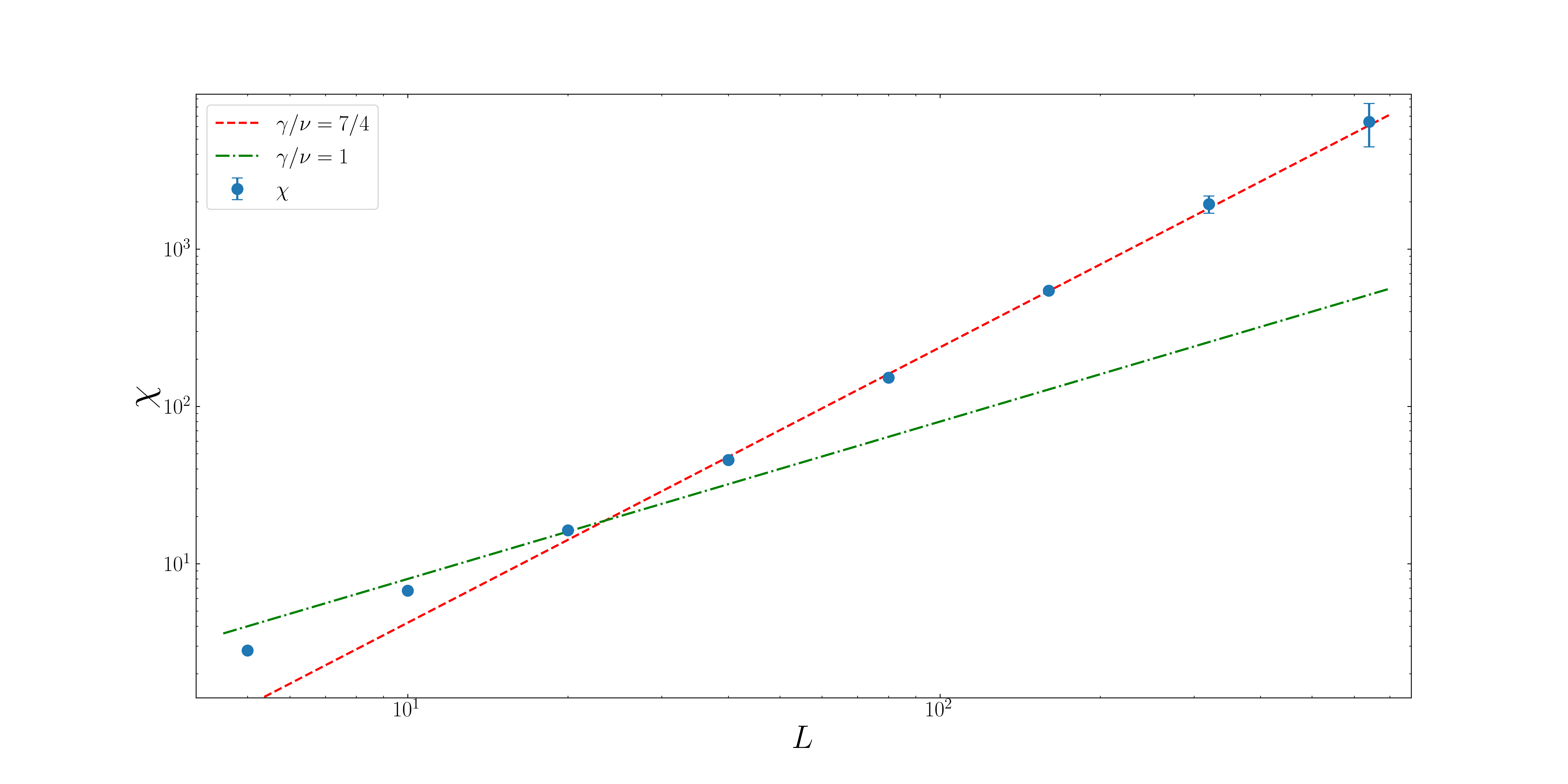}}
\caption{\flabel{off_lattice_m_and_chi} Finite size scaling of the magnetization $\langle|m|\rangle$, \Eref{def_m}, and the susceptibility $\chi$, \Eref{ture_susceptibility}, at the effective $\invTempCrit(L)$ for the off-lattice model with system sizes $L$ ranging from $10$ to $320$.}
\end{figure*}

\subsection{On-lattice particle model}\seclabel{on_lattice_model}
In the \emph{on-lattice model} spins $\sigma_\ivec$ are located at the vertices $\ivec\in\{1,2,\ldots,L\}^2$ of a $L\times L$ square lattice with periodic boundary conditions. The lattice is updated in two different types of sweeps, whereby spins $\ivec$ are updated sequentially as follows:

1) \emph{Local thermalisation}: Defining, similar to \Eref{def_hi_off_lattice}, 
\begin{equation}\elabel{def_hi_on_lattice}
h_\ivec^\pm = \sum_{\jvec.\text{nn}.\ivec}^N \delta_{\sigma_\jvec,\pm1}
\end{equation}
where $\jvec.\text{nn}.\ivec$ runs over all site indices $\jvec$ that are nearest neighbours of $\ivec$, every spin at $\ivec$ is updated to value $\sigma_\ivec=\pm1$ using the heat-bath algorithm, \ie with probability $\exp{\invTemp J h_\ivec^\pm}/(\exp{\invTemp J h_\ivec^+}+\exp{\invTemp J h_\ivec^-})$. The update of the spin orientation on the lattice thus follows that off the lattice, \cf \Eref{def_hi_off_lattice}. It is commonly assumed that it makes no noticeable difference to update in a random or a fixed sequential order \cite{newman_1999}.

2) \emph{Swapping}: Visiting every site $\ivec$ in a random sequential order, its state is swapped with that of one of its randomly selected nearest neighbours. Performing the swaps in a fixed order would introduce a bias whereby those sites that are updated early have a larger effective diffusion constant.

In each of these sweeps all spins are updated. To mimic on the lattice the off-lattice diffusion, \Sref{off_lattice_particle}, by characteristic distance $v_0$ in a time step, the nearest neighbour swapping is repeated $v_0^2$ times. 
After one thermalisation sweep, followed by $v_0^2$  swapping sweeps, time $t$ advances by one unit. \Fref{SIM_snapshot} shows the dynamical evolution of the quenching process in the on-lattice model.

\subsection{On-lattice field-theory}\seclabel{ModelB}

Without the swapping, the dynamics above is that of the Ising Model with an adjusted temperature, essentially a discretised version of \Eref{Hamiltonian_ModelC} with $g=0$ and the $\rho(\xvec,t)$ constantly that of a fixed lattice, or \Eref{def_Hamiltonian} with fixed $\xvec_i$.
The additional swapping noise may be thought of as a random current, as seen in Model B \cite{HohenbergHalperin_1977}. In the continuum limit, this is captured by 
\begin{align}
\elabel{eom_ModelA+Bnoise}
\dot{\phi}(\xvec,t) & = 
-\Big(
r \phi(\xvec,t) + \frac{u}{3!} \phi^3(\xvec,t)\Big) + \nu \nabla^2 \phi(\xvec,t) \nonumber \\
& \qquad + \xi(\xvec,t) + \eta(\xvec,t)
\end{align}
which is Model~A, \Erefs{eom_ModelC_phi} at $g=0$, with the conserved noise of Model B. 
\Eref{eom_ModelA+Bnoise} thus contains a non-conserved noise $\xi$ with noise correlator \Eref{ModelC_noise_correlators_xi} and a conserved noise $\eta$ with an amplitude, a length, of $v_0$,
\begin{equation}\elabel{conserved_noise_correlator}
\ave{\eta(\xvec,t)\eta(\xvec',t')} = - v_0^2 \nabla^2 \delta(\xvec-\xvec')\delta(t-t')
\ .
\end{equation}

The presence of the conserved noise renders the setup non-equilibrium, simply as the Fokker-Planck equation of \Eref{eom_ModelA+Bnoise} no longer allows for the stationary distribution of the form $\exp{-\HC/(\Gamma^2/2)}$, because of the presence of another noise amplitude. This noise amplitude carries a space dependence that does not cancel with that of the Hamiltonian as it happens in, say, Model~B, where $|\kvec|^2\HC$ in the numerator cancels with, say, $\half|\kvec^2|v_0^2$ in the denominator after Fourier-transforming.

However, this conserved noise is field-theoretically irrelevant compared to the non-conserved noise and hence can be ignored. Just like in Model~C, we therefore expect equilibrium thermodynamics to be restored on the large scale.

\section{Results}\seclabel{results}

\subsection{Off-lattice numerics}\seclabel{Off_lattice_result}
To investigate the validity of the droplet picture, \Sref{droplet_picture}, we have simulated the Brownian $q=2$ Potts Model systematically at increasing system sizes and large diffusion constants, with particle density $\rho =2$, interaction range $r_0=1$ and hopping length $v_0=2$.  
The density was chosen deliberately to exceed the disc percolation limit $\rho \approx 1.54(10)$ \cite{Mertens_2012}, so that dense clusters of particles with the same orientation form without the need of diffusion as a means of interaction, in the hope that we would observe how the diffusion undermines the ordering in dense clusters.

We simulated systems up to linear size $L=640$ with up to $10^7$ time steps, which is well beyond the time when the system reaches steady state. If the droplet picture applies, we expect the scaling behaviour of the model to change from standard Ising to, asymptotically, random-neighbour, \APref{randomNeighbourIsing}, at $L>\ell_\droplet$, which is when the system size is large enough to reveal that diffusion limits the formation of ordered clusters.
We expect another random-neighbour regime, namely when the system size $L$ is of the same order as $v_0$, so that all particles explore the whole system within a single sweep.

\begin{figure}[!t]
\centering
\includegraphics[width=1\columnwidth]{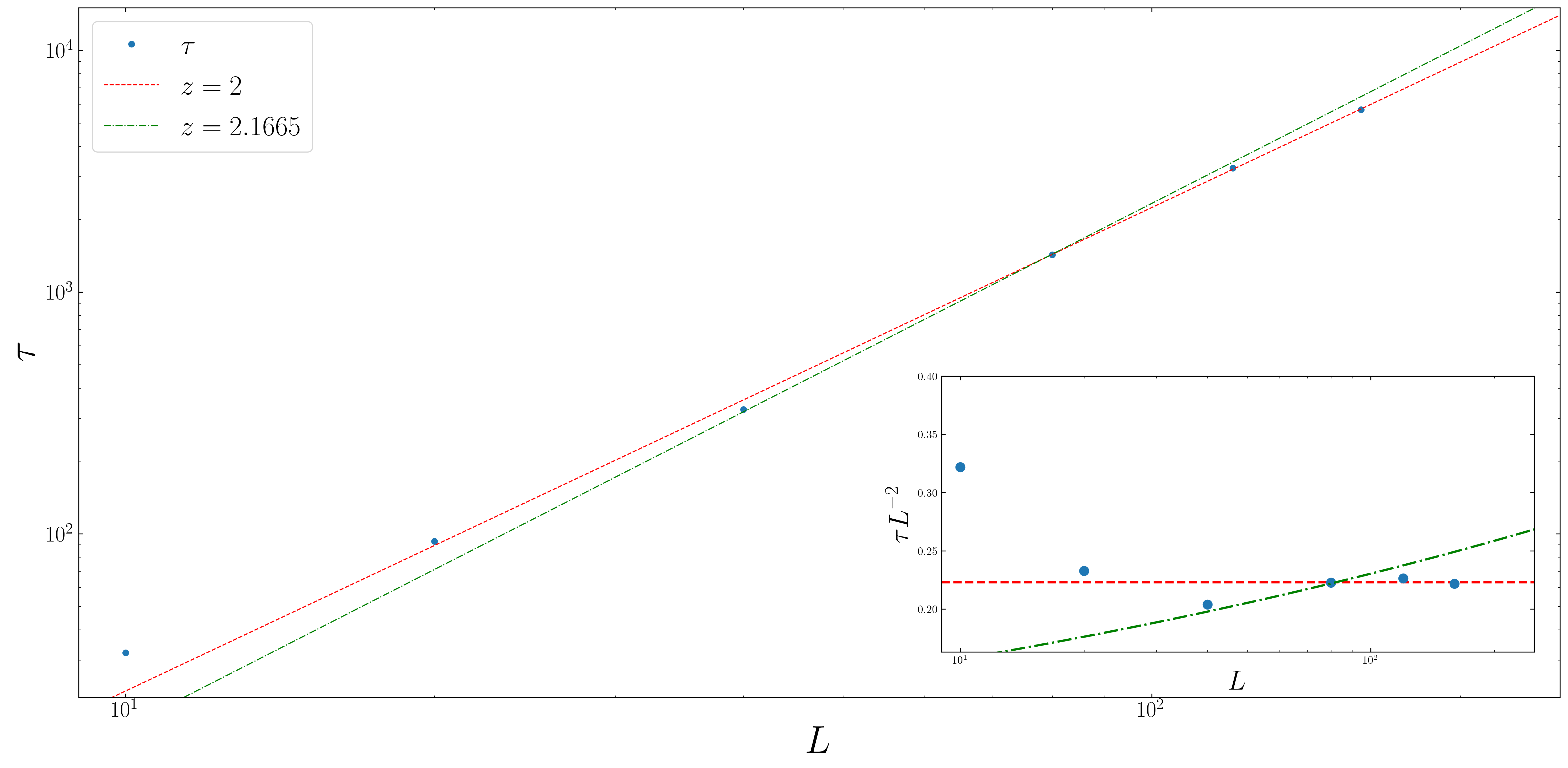}
\caption{\flabel{off_lattice_z} Finite size scaling of the correlation time $\tau$ for the off-lattice model with system sizes $L$ ranging from $10$ to $160$. Each data point represents an average of 500 independent simulations.}
\end{figure}

The data shown in \Fref{off_lattice_m_and_chi} confirms the existence of a crossover at small $L$ from a behaviour reminiscent of random-neighbour scaling to standard Ising scaling.
However, what is completely absent is the third, ``droplet regime'', when $L>\ell_\droplet$. 
We have studied large system sizes $L$, large densities $\rho$ and large effective diffusion constants $v_0$, in an effort to enter the regime where diffusion destroys ordered clusters beyond $\ell_\droplet$, but we have not found any signature of this phenomenon. 
In other words, the off-lattice $q=2$ Brownian Potts Model displays, as far as its static exponents are concerned, Ising behaviour for all large $L$ we considered. We relegate the discussion of finding no droplet regime to \Sref{no_droplet}.

On the other hand, we can confirm that the dynamical exponent in the present system, \Fref{off_lattice_z}, is not that of the standard Model~A, Glauber or heat-bath dynamics, of $z=2.1665(12)$ \cite{Nightingale_1996}, but the diffusive $z=2$.
As indicated above, \Sref{ModelC}, this can be explained with the dynamics of Model~C, even when the present setup is not equilibrium.

The dynamical exponent is difficult to extract, as the estimate $\corr(t)$, \Eref{def_twoTimeCorr} requires a significant amount of CPU-time to produce clean enough statistics to allow for a reliable estimate of the correlation time. We found that the dataset should be at least one hundred times the correlation time itself, so that $M$ in \Eref{def_twoTimeCorr} should be around $6\cdot10^5$ for the largest system sizes studied, \Fref{off_lattice_z}. It is difficult to discern $z=2$ from $z=2.1665$, best done by suitable rescaling as in the inset of \Fref{off_lattice_z}.
We observe $z=2$ asymptotically for the largest three system sizes $L=80,120,160$. It appears that smaller system sizes suffer from significant corrections up until a crossover, consistent with the one seen in \Fref{off_lattice_m_and_chi}.

The effect of the diffusion of the interaction partners on the scaling behaviour of the system at the phase transition, is thus quite subtle: The static exponents are the same as in Model~A, as if the diffusion poses no relevant perturbation, and yet the dynamical exponent reverts to the trivial, diffusive value. This is indeed the same mechanism as observed in Model~C. All exponents we find here for the off-lattice model are thus consistent with Model~C. Those are the standard static Ising exponents, as found in Model~A, plus a dynamical exponent of $z=2$.

\subsection{On-lattice numerics}
Off-lattice diffusion has two effects: Transport of spin-orientation and fluctuation in the number of interaction partners.
In order to differentiate the two, we devised the lattice model in \Sref{on_lattice_model} that allows for spin-diffusion without introducing fluctuations of the number of interaction partners. As outlined in \Sref{ModelB}, the additional \emph{conservative} fluctuations may be thought of as the same conserved noise as found in Model~B. Although this renders the setting out-of-equilibrium, the conserved noise is easily identified as field-theoretically irrelevant and hence should not have any effect on any of the exponents characterising the phase transition. The lattice model hence should amount to Model~A with irrelevant Model~B noise.

This is indeed confirmed by the numerics. \Fref{on_lattice_m_and_chi} shows the scaling of  $\ave{|m|}$ and $\chi$, confirming the standard static exponents of the Ising Model. \Fref{on_lattice_z} further shows the dynamical exponent $z=2.166$, compatible with Model~A $z=2.1665(12)$ \cite{Nightingale_1996}.  The largest system size we considered was $L=320$. As expected, the present setting recovers Model~A despite the presence of Model~B noise.

\begin{figure*}[!t]
\centering
\subfloat[\flabel{FSS_mag} Finite size scaling of the magnetization  $\ave{|m|}$.]
{\includegraphics[width=1\columnwidth]{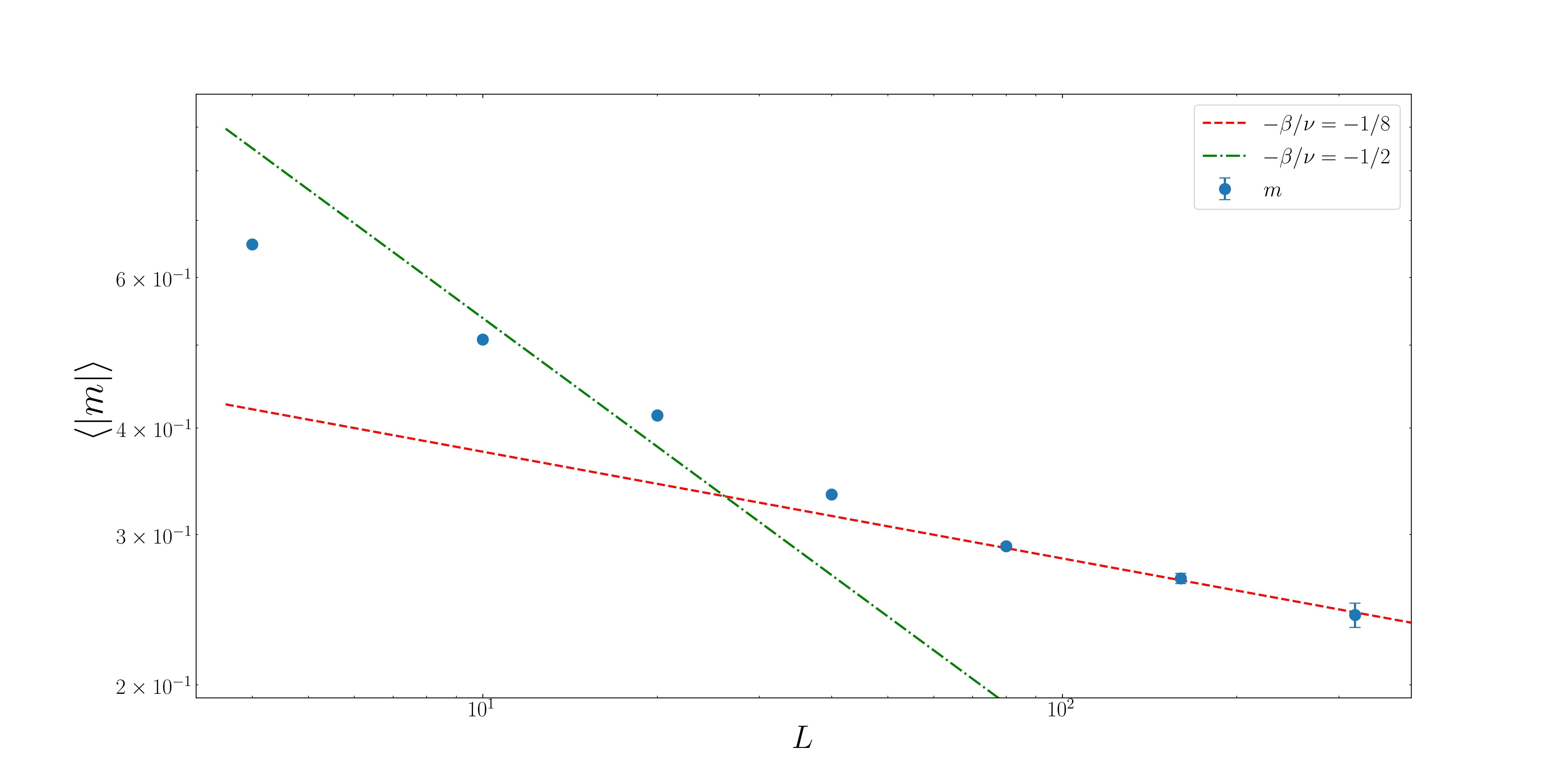}}
\subfloat[\flabel{FSS_chi} Finite size scaling of the susceptibility $\chi$. ]
{\includegraphics[width=1\columnwidth]{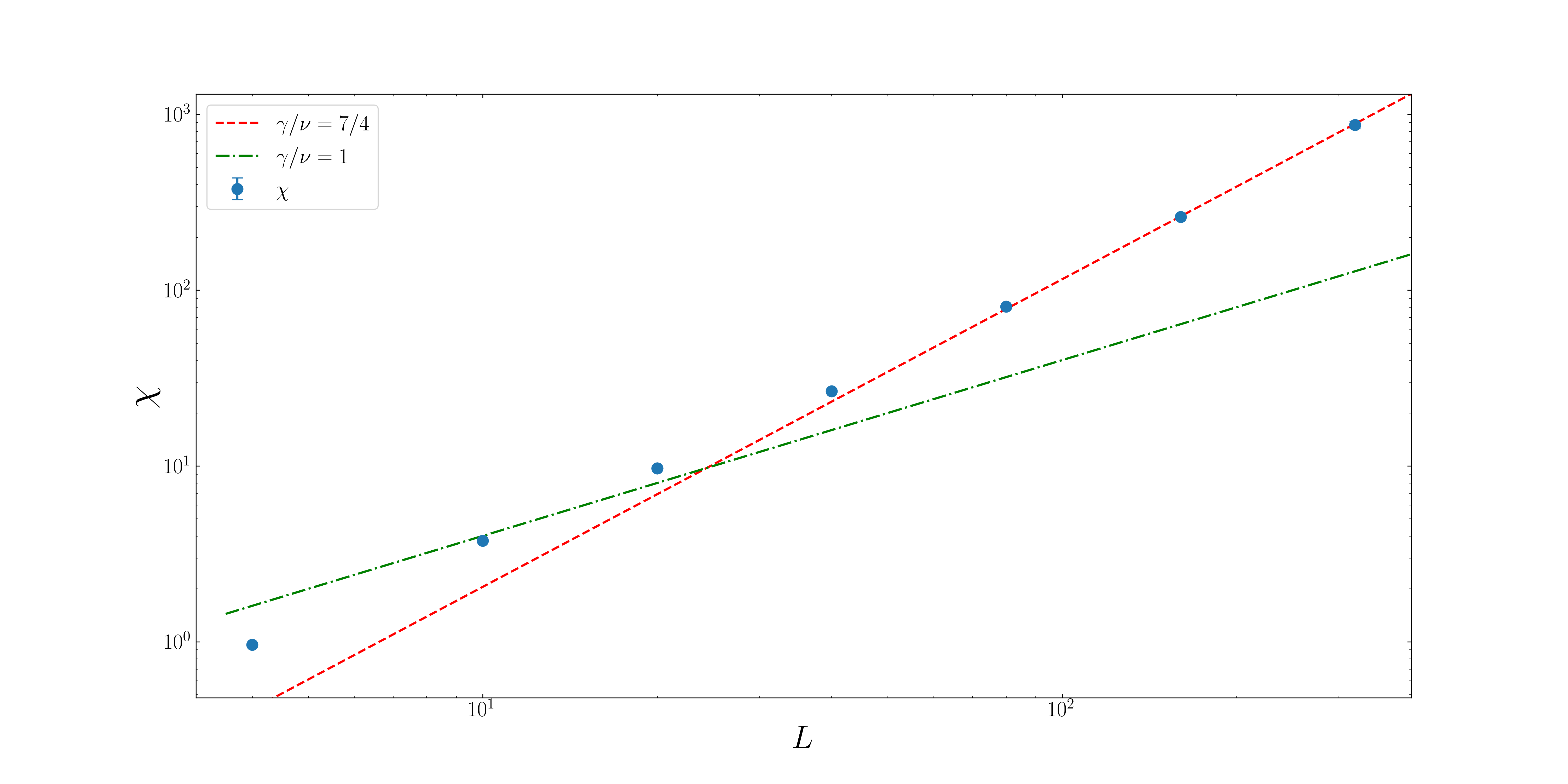}}
\caption{\flabel{on_lattice_m_and_chi} Finite size scaling of the magnetization  $\ave{|m|}$ and the susceptibility $\chi$ at $\invTempCrit$ for the on-lattice model with system sizes $L$ ranging from $10$ to $320$.}
\end{figure*}

\begin{figure}[!b]
\centering
\includegraphics[width=1\columnwidth]{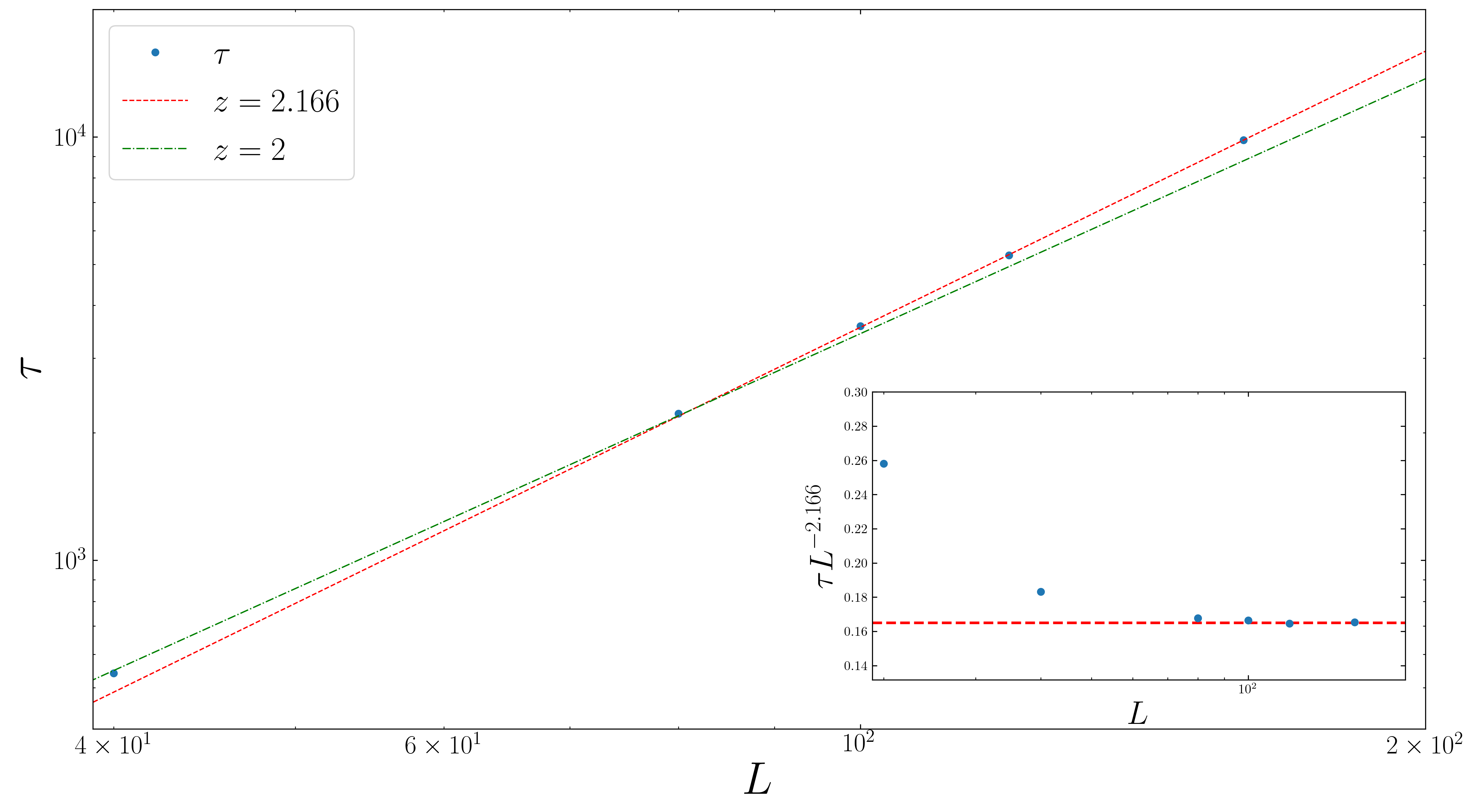}
\caption{\flabel{on_lattice_z} Finite size scaling of the correlation time $\tau$ for on-lattice model with system sizes $L$ ranging from $20$ to $160$. Each data point represents the average of 500 independent simulations.}
\end{figure}

\section{Discussion and Conclusion}\seclabel{discussion}
\subsection{The droplet picture}\seclabel{no_droplet}

The droplet picture in \cite{Woo_2022}, \Sref{droplet_picture}, suggests that ordered Potts (or Ising) clusters in the off-lattice model dissolve as oriented particles leave them by diffusion. This is being put forward as an explanation for the absence of phase coexistence in the Brownian $q>4$ Potts Model. 
However, this mechanism would have significant consequences for the $q=2$ Potts Model, at least at sufficiently large scales.
In the present work, we find no signal of the \emph{static} Ising scaling behaviour being lost or changed on large scales. We conclude that \emph{diffusion does not generally destroy ordered clusters}, but rather ``propagates spin order'' \cite[][p~7]{Woo_2022}.

We do find, as done previously \cite{Woo_2022}, that the dynamical exponent is changed from Model~A's $z=2.1665(12)$ \cite{Nightingale_1996}, to be expected without diffusion, to Model~C's $z=2$ \cite{HohenbergHalperin_1977}, meaning that the diffusion of ordered spins is a relevant perturbation to the dynamics, but irrelevant as far as the statics are concerned. We suggest that in the present setting \emph{diffusion spreads interaction} in the same way it does in the active Ising Model by \citet{Solon_2013,Solon_2015}. 

On the other hand, the lowering of the dynamical exponent to $z=2$ indicates a faster \emph{decay of correlations} in the presence of diffusion. 
One might therefore conclude that diffusion destroys correlations. However, increasing the diffusion from $v_0=2$ to $v_0=3$ \emph{lowers} the effective $\invTempCrit(L)$, \ie at higher diffusion constants the same setup orders at higher temperatures, \Fref{DIM_betac}. This is perfectly in line with the argument made above, namely that diffusion enhances interaction. In the same vein, diffusion accelerates the speed with which correlations decay in time. 
Irrespective of that, the statistics of the ordered clusters, as characterised by the static exponents, remains unchanged, \ie the diffusion has no deleterious effect on static correlations.
It seems that, on large enough scales $L$, any ``diffusion length'' $v_0$ will produce Model~C behaviour, rather than dissolving clusters.

The hypothesis of diffusion-mediated interaction may best be tested in the dilute limit, where particles interact primarily in a delayed fashion by diffusing into each other, as opposed to the dense limit, where particles are interacting directly and immediately as they are packed into dense clusters with excitations capable of sweeping the vast majority of the system. The dilute limit is in fact the density chosen in \cite{Woo_2022}, as the disc percolation threshold in the continuum is $1.54(10)$ \cite{Mertens_2012}, clearly greater than $1$ in \cite{Woo_2022} and clearly less than $2$ chosen in our setup.

In \Fref{off_lattice_z} we illustrate our finding of $z=2$ from the correlation function \Eref{def_twoTimeCorr}, in line with \citeauthor{Woo_2022}'s finding of what is called $Z_C$ in their work. In addition to this exponent, the authors further found another dynamical exponent $Z_R=2.15(10)$ as the exponent governing the spreading of correlations starting from the time of the system being quenched from an infinite temperature to the critical temperature. As the present model is generically out of equilibrium, the presence of another dynamical exponent in a very similar observable may not surprise.

We can confirm similar findings for the short time critical behaviour \cite{Lzb_1995,Zheng_1998} of the onset of a finite ensemble averaged order parameter for a system starting from a completely ordered state, \ie, $m(t=0)$=1, 
the Binder cumulant $U(t)\equiv \ave{m^2(t)}/\ave{|m(t)|}^2 -1 \propto t^{d/z}$ \cite{zheng_2002}, using $d=2$. The exponent $Z_R\approx 2.15$ being larger than the diffusive exponent of $z=2$ suggests that the establishing of order in the system is not governed by diffusion, but by a slower dynamics, namely one that takes longer, $\ell^{2.15}$ \vs $\ell^{2}$, to reach the same distance $\ell$. On the other hand, the faster, diffusive dynamics does govern the decay of correlations.

We may thus speculate that the diffusion governs the local propagation of correlations without advancing order or disorder, while the global onset of order is governed by the slower dynamics of Model~A. This \emph{global} ordering dynamics does not seem to benefit from the diffusion process whose r{\^o}le in spreading order must eventually peter out as the diffusing spins flip orientation, for example when they are no longer surrounded by any neighbours.

\begin{figure}[!t]
\centering
\includegraphics[width=1\columnwidth]{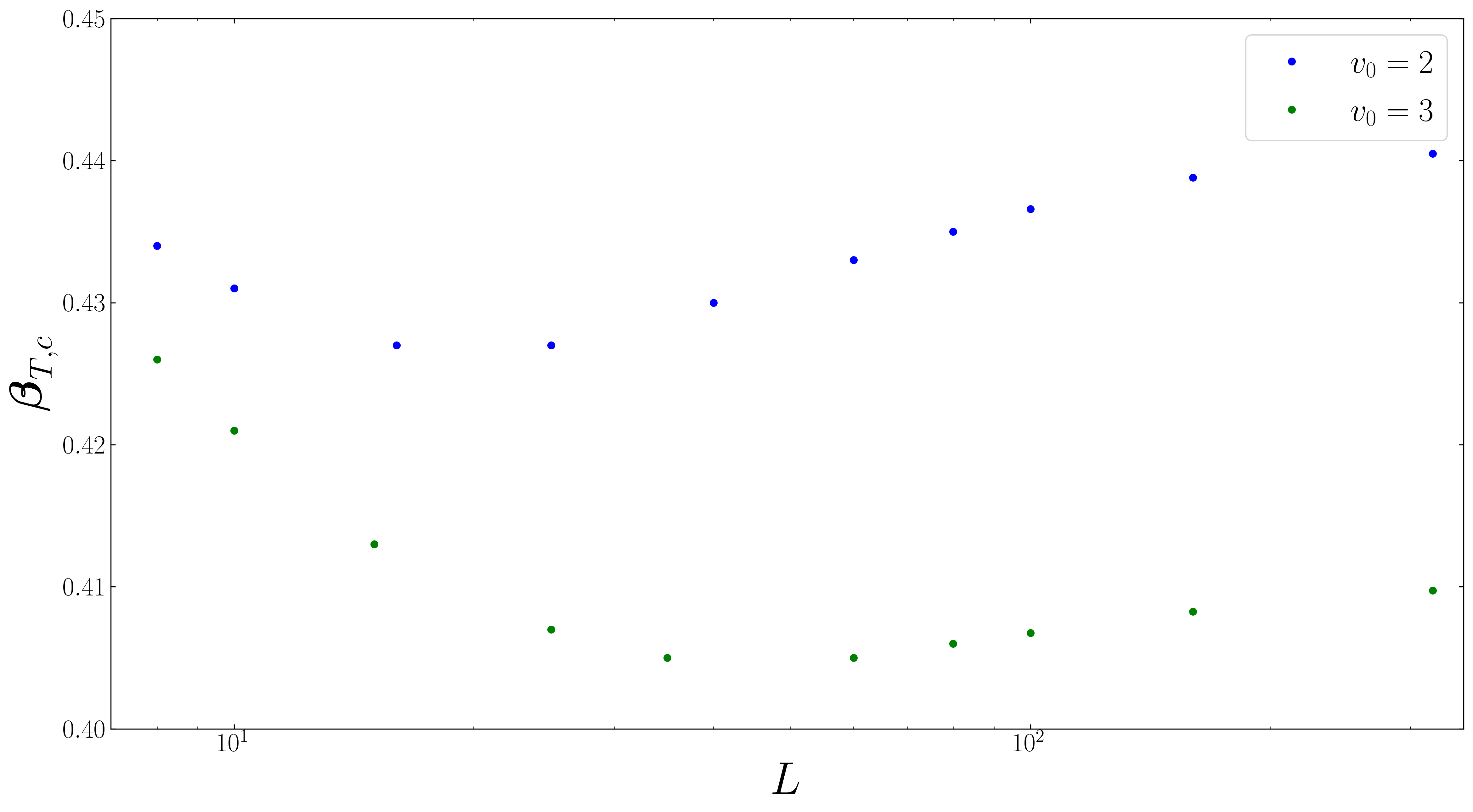}
\caption{\flabel{DIM_betac} Effective critical inverse temperature $\invTempCrit(L)$, as determined by the maximum of $\chi$, \Sref{observables}, for off-lattice model with different $v_0$. Larger $v_0$ leads to ordering at higher temperatures, lower $\invTempCrit$.}
\end{figure}

\subsection{Diagrammatics}\seclabel{diagrammatics_ModelC}
The Hamiltonian $\HC_C$  of Model~C, \Eref{Hamiltonian_ModelC}, is an effective theory, based primarily on the hydrodynamic assumption that there is a global field $\phi(\xvec,t)$ that permeates all space and that provides particles with their internal degree of freedom given one is present. There is no notion in its equation of motion \Eref{ModelC_Langevin_from_Hamiltonian_phi} of having too low a density or, in fact, no particles at all. It is written down under the assumption of there being enough of a background density to sustain global excitations, while $\rho(\xvec,t)$ captures only the fluctuations about this background. This way, the two degrees of freedom are separated out into two variables, which can evolve in very standard ways, even though $\phi(\xvec,t)$ really ``ought to know more'' about $\rho(\xvec,t)$. The situation is comparable with the velocity field evolving in the Toner-Tu equations \cite{TonerTu:1995} rather than the current, which, in the absence of particles, ought to vanish in most of space.

Even with these allowances in place, $\HC_C$ in \Eref{Hamiltonian_ModelC} does not capture correctly the density fluctuations as encountered in the off-lattice model, as they should enter via the (real-space) interaction term. After all, it is particle \emph{interaction} that is affected by the diffusion of interaction partners. 
In other words, the Langevin equations to study are
\begin{subequations}
\elabel{eom_preModelC}
\begin{align}
\dot{\phi}(\xvec,t) & = 
-\Big(
r \phi(\xvec,t) + \frac{u}{3!} \phi^3(\xvec,t)
\Big) 
\elabel{eom_preModelC_phi}
\\
\nonumber
& \qquad + \big(\nu + \mu \rho(\xvec,t) \big) \nabla^2 \phi(\xvec,t) 
+ \xi(\xvec,t)\\
\elabel{eom_preModelC_rho}
\dot{\rho}(\xvec,t) & = \nabla^2 \transDiffusion  \rho(\xvec,t) 
+ \zeta(\xvec,t)
\end{align}
\end{subequations}
rather than \Erefs{eom_ModelC}. Diagrammatically \cite{Tauber_2008}, the $\mu$-vertex in \Eref{eom_preModelC_phi}, which is in fact field-theoretically irrelevant, corresponds to 
\begin{equation}\elabel{ActualVertex}
    \begin{tikzpicture}[baseline=(current  bounding  box.center)]
    \node at (0,0) {\includegraphics[width=1.8cm]{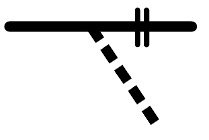}};
    \end{tikzpicture}
\end{equation}
where the full lines represent the (amputated) response propagator of $\phi$ and the dashed line the (amputated) response propagator of $\rho$. The double dash on the incoming response propagator indicates the Laplacian in $\mu\rho\nabla^2\phi$ of \Eref{eom_preModelC_phi}. This is in contrast to Model~C's original $g$-vertex of \Eref{eom_ModelC_phi}
\begin{equation}\elabel{ModelCVertex}
    \begin{tikzpicture}[baseline=(current  bounding  box.center)]
    \node at (0,0) {\includegraphics[width=1.8cm]{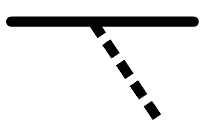}};
    \end{tikzpicture} \ .
\end{equation}
However, this vertex, \Eref{ModelCVertex}, is generated from \Eref{ActualVertex} via the $\phi^4$-vertex and the noise through the diagram, 
\begin{equation}\elabel{ModelCFromActual}
    \begin{tikzpicture}[baseline=(current  bounding  box.center)]
    \node at (0,0) {\includegraphics[width=2.3cm]{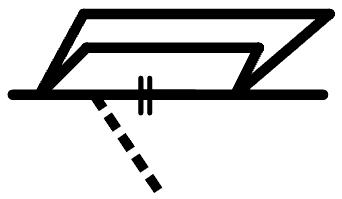}};
    \end{tikzpicture} \ ,
\end{equation}
so that \Eref{eom_preModelC} effectively, on large scales, ends up being Model~C, \Erefs{eom_ModelC},  with $w=0$ in \Erefs{eom_ModelC_rho}. However, such a non-equilibrium modification of Model~C is known \cite{Akkineni_2004} to still result, ultimately, in equilibrium Model~C behaviour \cite{ScandoloPauschCates_2023}. This requires further qualification: Setting $w=0$ in the present work amounts to $\Theta=0$ in \cite{Akkineni_2004} (which is parameterised altogether differently and contains a parameter $w$ as well, however, with a different meaning), which in their Sec.~III.E.2 is found to be an unstable fixed point. As opposed to our $w$, which captures the mechanical feedback of the spin orientation to the position and thus cannot be generated, $\Theta$ in \cite{Akkineni_2004} captures the ratio of the noise amplitudes \Eref{ModelC_noise_correlators} and may be generated by otherwise irrelevant couplings, allowing the system to escape from bare $\Theta=0$ and slipping into standard Model~C behaviour. Even if that does not happen, $\Theta=0$ seems to result in $z=2$ and Model~A's (and Model~C's) static exponents regardless. It is beyond the scope of the present work, to show directly, for our setup with $w=0$ in \Eref{eom_ModelC_rho}, that Model~C exponents are produced.

\subsection{The Harris criterion}

Above we are making the case for the diffusion not having the effect described in the droplet picture, \Sref{droplet_picture}, which is being invoked in \cite{Woo_2022} to explain the phase transition in the $q=5,6,7,8$ Potts Model being \emph{continuous} in the presence of diffusion, in contrast to the transition being discontinuous for $q>4$ without diffusion \cite{Yeomans_book_1992}. While the authors find that the exponents for the $q=2$ case, which is the focus of the present work, are not or barely affected by the diffusion, for $q=3$ the exponent $1/\nu$ is found to be $1.03(10)$ instead of $6/5$ and for $q=4$ similarly $1.05(10)$ instead of $3/2$ \cite{Yeomans_book_1992}. 

This finding is in line with the Harris criterion \cite{Harris_1974,Brooks_2016}, which suggests that disorder affects critical behaviour when the width of fluctuations, say $\sigma(T_{c,\text{\tiny eff}})$ of an effective, local, critical temperature exceeds the distance to the supposed global critical point, $|T-T_c|$. The width of the fluctuations is determined by the volume the disorder is averaged over, so that it scales with the correlation length $\xi$ like $\xi^{-d/2}\propto |T-T_c|^{\nu d/2}$. Considering small $|T-T_c|$ the disorder is expected to be relevant if $\nu d/2<1$, which, translates to $\alpha>0$ using standard scaling relations \cite{Stanley_book_1971}.

The Ising Model and thus the $q=2$ Potts Model have $\alpha=0$ and therefore represent a marginal case. For $q=3,4$ the exponents are $\alpha=1/3$ and $\alpha=2/3$ respectively \cite{Yeomans_book_1992,Wu_1982}, so disorder is expected affect the critical behaviour \cite{Olson_1999}. 
For $q>4$, disorder is known to affect the nature of the transition \cite{Rafael_2010,Aizenman_1989,Cardy_1997,Olson_1999,Imry_1979,Hui_1989}. In fact, rounding of a first order phase transition has been observed generic in the art of equilibrium in the presence of quenched random fields \cite{MartinBonachellaMunoz:2014}.

We conclude that the change of the critical behaviour for the $q>2$ Potts Model is well explained by the Harris criterion and in fact established, and that the $q=2$ Potts Model's unchanged behaviour is in fact in line with this explanation.

\subsection{On-lattice Model}\seclabel{on_lattice_conclusions}
The on-lattice model studied in \Sref{on_lattice_model} allowed us to determine whether the different dynamical exponent, $z=2$ rather than $z=2.1665(12)$, in the off-lattice model is due to the randomisation of the number of interaction partners, absent on-lattice, or due the diffusive mediation of interaction, present on-lattice. As it turns out, the diffusive ``transport of spin orientation'' on the lattice affects neither the static nor the dynamical exponents of the $q=2$ Potts Model, indicating that the dynamical exponent observed in the off-lattice model is due to the randomisation of the number of interaction partners. As the droplet picture would equally apply on-lattice, the absence of any effect of the diffusion confirms once more that the droplet picture does not apply and that the effect of diffusion is merely an additional form of interaction, that on-lattice has no effect on the critical behaviour, and off-lattice affects only the dynamical exponent.

\subsection{Overall conclusion}
The critical behaviour of the Brownian $q=2$ Potts Model or, equally, the Brownian Ising Model can be well understood on the basis of well-established field theoretic models. There is no need to invoke additional mechanisms, such as the droplet picture, which do not seem to produce a consistent picture. We have performed extensive numerics to confirm this. 

Beyond the scope of the present work are the $q>2$ Potts Models, whose critical behaviour, however, can be well understood on the basis of the Harris criterion. It remains to be confirmed whether this is consistent with numerical simulations of the $q>2$ Brownian Potts Model at larger system sizes and densities.

\section*{Acknowledgements}
The authors would like to thank 
Ron Dickman,
Henrik Jensen, 
Jacob Knight
and
Emir Sezik for helpful discussions and suggestions, and Andy Thomas for providing invaluable technical support.

\bibliography{articles}

\appendix
\begin{widetext}
\section{Random neighbour Ising Model}\seclabel{randomNeighbourIsing}

The random neighbour version of the Ising Model discussed in the following captures the $v_0\to\infty$ limit of the on-lattice model introduced in \Sref{on_lattice_model}. 
By extension, \APref{Off_lattice_random}, it further captures the off-lattice model as well, \Sref{off_lattice_particle}. To make the correspondence exact, the updating of every spin degree of freedom needs to be done at random times, notably only ever after an update of the spatial degrees of freedom. 

Concretely, the random neighbour model is defined as follows: Spins $\sigma_\ivec\in\{-1,1\}$ are indexed by $\ivec\in\{1,2,\ldots,L\}^2$ on an $L\times L$ square lattice. They are updated by randomly shuffling them across the lattice and then applying a single heat bath update at a single, randomly chosen site. Alternatively, one might select an initial site at random, choose its four distinct nearest neighbours at random from the other sites and finally update the initial site. In fact, because the spatial layout plays no r{\^o}le in the updating, one may think of a pool of $N^+$ sites with spin $+1$ (up) and $N^-$ sites with spin $-1$ (down), such that $N^++N^-=N=L^2$. To update, a spin is chosen at random, as are $q=4$ ``nearest neighbours'', and the state of the spin is updated according to the heat bath algorithm, which assigns the new state solely on the basis of the number of neighbouring up versus down spins. To have access to a well-defined susceptibility, we also allow for an external field $H$. Defining now, similar to \Eref{def_hi_on_lattice},
\begin{equation}\elabel{def_delta_H}
\half \delta \HC(n^+) = -J\big(n^+-(q-n^+)\big)-H
\end{equation}
as the contribution to the overall energy when the spin with $n^+$ up neighbours and $n^-=q-n^+$ down neighbours is oriented up, a heat bath update results in up with probability 
\begin{equation}\elabel{def_pplus}
p^+(n^+;\alpha,H)
=\frac{\exp{-\beta\delta\HC}}{\exp{-\beta\delta\HC}+\exp{\beta\delta\HC}}
=\frac{\exp{\alpha(2n^+-q)+2\beta H}}{2\cosh\big(\alpha(2n^+-q)+2\beta H\big)}
\end{equation}
where $\alpha=2\beta J$ and $\delta\HC$ is evaluated at $n^+$. Otherwise the outcome is a down spin.

\Eref{def_delta_H} implements Ising spins, rather than Potts states, which interact in the form $\delta_{\sigma_i,\sigma_j}=(1+\sigma_i\sigma_j)/2$, \Eref{Potts_vs_Ising}, and which covered in the current derivation by replacing $J$ above or $\alpha$ below, as well as $H$, by half their value, \ie as if the temperature was twice as high. In other words, the Potts Model with interaction strength $J$, external field $H$ and inverse temperature $\beta$ is implemented by dividing $\alpha$ and $H$ used in the following by $2$. On the lattice, where the number of interaction partners $q$ is fixed, the offset by $1/2$ amounts to a constant, overall shift in the energy, independent of the spins and thus not modifying $\delta \HC(n^+)$. Off the lattice, \Sref{off_lattice_particle}, the number of interaction partners varies, so that the overall shift depends on the random number of neighbours. However, this random contribution to the Hamiltonian does not enter into $\delta \HC(n^+)$, nor does it bias in any form the random number of neighbours, so that the offset does not enter at all into any of the dynamics.

If the probability to select $n^+$ up spins and $n^-=q-n^+$ down spins is $C_{n^+}(N^+)$, the master equation for $\PC(N^+,t)$, the probability to have a total of $N^+$ up spins and thus $N^-=N-N^+$ down spins after $t$ updates reads
\begin{multline}
\PC(N^+, t+1) = 
\frac{N^+ + 1}{N}
\PC(N^+ + 1,t)
\sum_{i=0}^q (1-p^+(i;\alpha,H)) C_{i}(N^+ +1)
+
\frac{N^+}{N}
\PC(N^+,t)
\sum_{i=0}^q p^+(i;\alpha,H) C_{i}(N^+)
\\
+
\frac{N-N^+}{N}
\PC(N^+,t)
\sum_{i=0}^q (1-p^+(i;\alpha,H)) C_{i}(N^+)
+
\frac{N-(N^+-1)}{N}
\PC(N^+ - 1,t)
\sum_{i=0}^q p^+(i;\alpha,H) C_{i}(N^+ -1)
\end{multline}
where the transitions are, in order of appearance, pick `up' and update to `down', pick `up' and update to `up', pick `down' and update to `down', pick `down' and update to `up'. 
The transition probabilities of the two updates that leave the state unchanged sum to unity, $\sum_i C_i(N^+)=1$, so that, more symmetric,
\begin{multline}\elabel{ME_V2}
\PC(N^+, t+1) - \PC(N^+,t)= \\
\left(
\frac{N^++1}{N}
\PC(N^++1,t)
\sum_{i=0}^q (1-p^+(i;\alpha,H)) C_{i}(N^+ +1)
-
\frac{N^+}{N}
\PC(N^+,t)
\sum_{i=0}^q (1-p^+(i;\alpha,H)) C_{i}(N^+)
\right)
\\
-\left(
\frac{N-N^+}{N}
\PC(N^+,t)
\sum_{i=0}^q p^+(i;\alpha,H) C_{i}(N^+)
-
\frac{N-(N^+-1)}{N}
\PC(N^+-1,t)
\sum_{i=0}^q p^+(i;\alpha,H) C_{i}(N^+-1)
\right)\ .
\end{multline}
The terms on the right are all the contributions to a change in the probability $\PC(N^+,t)$ to have $N^+$ up spins. In order of appearance, they are the positive influx from the state with $N^++1$ up spin undergoing a down-flip, resulting in $N^+$ up spins, and similarly loss due to $N^+\to N^+-1$, loss due to $N^+\to N^++1$ and gain due $N^+-1\to N^+$.

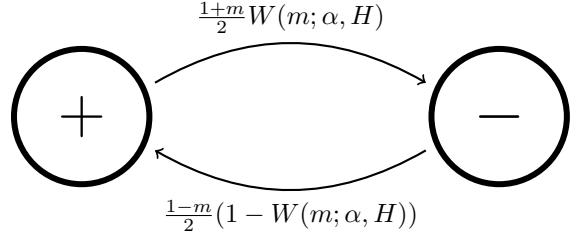
\begin{figure}
    \centering
    \begin{tikzpicture}[scale=1.5]
        \draw[line width = 0.8mm] (-1.85,0) circle (0.6cm) node {\Huge $+$};
        \draw[line width = 0.8mm] (1.85,0) circle (0.6cm) node {\Huge $-$};
        \path[thick, ->] (-1.2,0.3) edge[bend left=30] 
            node[midway,above] {$\frac{1+m}{2} W(m;\alpha,H)$} (1.2,0.3);
        \path[thick,->] (1.2,-0.3) edge[bend left=30] 
            node[midway,below] {$\frac{1-m}{2} (1-W(m;\alpha,H))$} (-1.2,-0.3);
    \end{tikzpicture}
    \caption{Cartoon of the updating of the populations of up and down spins under the random neighbour dynamics, using the transition probability defined in \Eref{def_W}, so that the labels of the arrows indicate the flux in and out of the state.}
    \flabel{m_flux_caricature}
\end{figure}

Demanding stationarity amounts to setting the left-hand side of \Eref{ME_V2} equal to zero. However, to determine $\PC(N^+,t)$ at stationarity requires implementing boundary conditions, as \Eref{ME_V2} applies only for $N^+\in\{0,1,\ldots,N\}$: Only with those boundary conditions, it is clear that $\PC(N^+,t)$ has no stationary solutions with a finite, constant current, as \Eref{ME_V2} is quasi-local in $N^+$ and therefore cannot incorporate the constraint of no net-flux across $N^+$. A more direct route to the properties of the stationary state is to consider the fluxes in and out of a given state given by $N^+$, or, equivalently by the magnetisation
\begin{equation}
    m = \frac{N^+ - N^-}{N}=\frac{2N^+}{N}-1
\end{equation}
such that $N^+/N=(1+m)/2$ and $N^-/N=(1-m)/2$. As depicted in \Fref{m_flux_caricature}, the magnetisation $m(t)$ evolves according to the updating ``rules'' outlined above, approximated by  
\begin{equation}\elabel{eom_m}
m(t+1)-m(t)
    =
    - \frac{2}{N} j(m; \alpha, H) + \xi(t)
\end{equation}
with continuous $m(t)\in\Rset$, rather than $m\in\frac{1}{N}\{-N,-N+2,\ldots,N-2,N\}$, and deterministic flux
\begin{equation}\elabel{def_j}
    j(m; \alpha, H) = 
    \half (1+m) \sum_{i=0}^q (1-p^+(i;\alpha,H)) C_{i}(N^+)
    - \half (1-m) \sum_{i=0}^q p^+(i;\alpha,H) C_{i}(N^+)
    \ ,
\end{equation}
which is the net flow of spin-up particles out of the left reservoir shown in \Fref{m_flux_caricature}, with each particle carrying away $2/N$ of the magnetisation from $m$. The noise $\xi$ shall be approximated by a Gaussian with vanishing mean and correlator
\begin{equation}
\ave{\xi(t)\xi(t')}=
\frac{2}{N^2}\delta_{tt'}
    \ ,
\end{equation}
chosen as to provide the correct fluctuations when $m=0$, in which case half the updates will result in a spin-flip, changing $m$ by $2/N$, producing a variance of $(1/2)(2/N)^2$.

To make further progress we define
\begin{equation}\elabel{def_W}
W(m;\alpha,H) = \sum_{i=0}^q \Big(1-p^+(i;\alpha,H)\Big) C_{i}(N^+)
\end{equation}
for the transition probability from up to down. 
The transition probability from down to up is identical up to a sign change of $H$ and $\alpha$, by inspection of \Eref{def_pplus},
$1-p^+(i;\alpha,H)=p^+(i;-\alpha,-H)$, so that 
\begin{equation}\elabel{1MW_in_W}
1-W(m;\alpha,H) = \sum_{i=0}^q p^+(i;\alpha,H) C_{i}(N^+) = W(m;-\alpha,-H)\ .
\end{equation}
Further, using that 
$C_{i}(N^+)=C_{q-i}(N-N^+)$ and 
$p^+(q-i;\alpha,H)=p^+(i;-\alpha,H)$,
one can show from \Eref{def_W},
\begin{align}
W(-m;-\alpha,H)
&=\sum_{i=0}^q (1-p^+(i;-\alpha, H)) C_{i}(N-N^+) \\
&=\sum_{i=0}^q (1-p^+(q-i;\alpha,H)) C_{q-i}(N^+) \\
&=\sum_{i=0}^q (1-p^+(i;\alpha,H)) C_{i}(N^+) \\
&=W(m;\alpha,H)
\ ,
\end{align} 
so that \Eref{1MW_in_W} may be furtrher rewritten as $\sum_{i=0}^q p^+(i;\alpha,H) C_{i}(N^+) = W(-m;\alpha,-H)$.

The flux \Eref{def_j} can now be written as 
\begin{equation}\elabel{j_clean}
    j(m; \alpha, H) = \half (1+m) 
    W(m;\alpha,H) 
    -  \half (1-m) \big(1-W(m;\alpha,H) \big)
    =
    \half (1+m) W(m;\alpha,H)
    - 
    \half (1-m) W(-m;\alpha,-H)
\ .
\end{equation}

\newcommand{\mbar}{\overline{m}}
At $H=0$ for any finite $N$ the expectation of $m$, the net magnetisation, has to vanish by symmetry. To determine a meaningful order parameter, we instead determine \emph{stable roots of the flux}, \ie values $\mbar$ of $m$ such that $j(\mbar;\alpha,H)=0$ in a way such that $m(t)$ as of \Eref{eom_m} will evolve back to $\mbar$ when perturbed away from it. 
Because the chain of states $\ldots,N^+-1,N^+,N^++1,\ldots$ at stationarity is bound to have vanishing net-flux between any two consecutive states, simply because there is no loop between any three or more states, the present system is bound to obey detailed balance at stationarity and thus is in equilibrium.

In the following, we will focus at first on $H=0$ and, as will be explained further below, $N\to\infty$, later also determining the finite size scaling of $\mbar$.
By inspection of \Eref{eom_m}, it is clear that stability requires $\partial_m j|_{m=\mbar}>0$, as to restore $m(t)=\mbar$ when $m(t)$ deviates from it. Further, according to the rightmost expression in \Eref{j_clean} the flux between states always vanishes when $H=0$ and $m=0$. By the first equality of \Eref{j_clean}, the flux vanishes for every $\mbar$ such that
\begin{equation}
\elabel{nontrivial_roots}
    \mbar = 1 - 2 W(\mbar;\alpha,H) \ ,
\end{equation}
which captures the trivial solution $\mbar=0$ because $W(0;\alpha,0)=1/2$ from $1-W(m;\alpha,H)=W(-m;\alpha,-H)$. At $H=0$, this latter identity guarantees that $1-2W$ is odd in $m$.

The non-trivial solutions of \Eref{nontrivial_roots} are the stable ones, $\partial_m j >0$, when $m=0$ loses stability, provided $\partial^3_m|_{m=0} W(m; \alpha, H)=:W^{(3)}(0;\alpha,H)>0$. To see this, we assume $m=0$ is not a stable solution of 
$j(m; \alpha, H)=0$ rewritten as
\begin{equation}\elabel{j_useful}
j(m; \alpha, H)=\half m - \half(1 - 2W(m; \alpha, H))\ ,
\end{equation}
using \Eref{j_clean}, so that $\partial_m|_{m=0} j <0$ and thus $\partial_m|_{m=0} W(m; \alpha, H)=-\half(1 + \epsilon)<-1/2$, with some (small) $\epsilon>0$. Expanding the right-hand side of \Eref{nontrivial_roots} in small $m$ at $H=0$ gives $\mbar= \mbar (1 + \epsilon) - (1/3) \mbar^3 W^{(3)}(0;\alpha,0) + \OC(\mbar^5)$, which has non-trivial roots 
\begin{equation}\elabel{m_solns}
\mbar=\sqrt{\frac{3\epsilon}{W^{(3)}(0;\alpha,0)}} + \hot
\end{equation}

The odd $1-2W$ in \Eref{j_useful} is easily expanded to give $-2\partial_m W = 1+ \epsilon -m^2 W^{(3)}$ and \Eref{j_useful} then produces $\partial_m j=\epsilon>0$ for the derivative of the current at the non-trivial solution \Eref{m_solns}. In other words, the non-trivial solution is stable.
At $H=0$, the current setup thus has the potential to display a phase-transition, from $\mbar=0$ to $\mbar\ne0$ namely precisely when $\partial_m|_{m=0}j(m;\alpha,0)$ changes sign. Assuming that this happens to leading order linear in $\alpha$ implies a linear dependence of $\epsilon$ on $\alpha$, say $\epsilon\propto\alpha-\alpha_c$, and therefore, \Eref{m_solns}, exponent $\beta=1/2$ as usual in mean-field theories \cite{Stanley_book_1971}. In \APref{concrete_W} we calculate $W(m;\alpha,H)$ explicitly and confirm $W^{(3)}(0;\alpha,0)>0$ and $\partial_m j>0$ for $\mbar$ according to \Eref{m_solns}.

At the assumed critical point $\alpha=\alpha_c$, the flux vanishes, $\partial_m|_{m=0}j=0$ and therefore $\partial_m|_{m=0} W(m; \alpha, 0)=-\half$ so that expanding the right-hand side of \Eref{nontrivial_roots}  in small $m$ and $H$ gives 
\begin{equation}
    \mbar=\mbar-(1/3) \mbar^3 W^{(3)}(0;\alpha,0) - 2 
    H \partial_H|_{H=0} W(0;\alpha,H) + \OC(H^2) 
    \ ,
\end{equation}
implying exponent $\delta=3$. All other, standard mean-field exponents follow from thermodynamics \cite{Stanley_book_1971}, given that the present system is in equilibrium. In addition, it can be analysed for its finite size behaviour, which we discuss next. 

In the preceding analysis, we have implicitly assumed that $N\to\infty$ when we suggested that $\mbar\ne0$ is possibly observed as the stationary state simply because this particular value of $m=\mbar$ renders the flux $0$ and excursions of $m$ away from $\mbar$ result in a restoring flux. At finite $N$, however, $m$, governed by \Eref{eom_m}, will escape from any $\mbar$ sooner or later and will thus arrive at $-\mbar$, resulting in vanishing average $m$. Without any current at work, it takes $\OC(N)$ updates in the ``opposite'' direction to flip $m$ from $\mbar$ to $-\mbar$ and thus typically $\OC(N^2)$ random updates to accumulate a fluctuation of order $N$. The effect of the current over the course of time $N^2$ is of order $N$, as it is bounded away from $0$ for most of the journey and enters with a prefactor of $2/N$ in \Eref{eom_m}. In other words, as $N$ increases, $m$ is increasingly trapped at $\mbar$ or $-\mbar$.

To quantify this, we analyse the continuous-time approximation of \Eref{eom_m},
\begin{equation}
    \dot{m}(t) = - \frac{2}{N} j(m;\alpha,H) + \xi(t)
\end{equation}
for which we can immediately write down the stationary distribution
\begin{equation}\elabel{stat_distribution_P}
    \PC_0(m;\alpha,H) \propto \Exp{-\frac{(2/N)V(m;\alpha,H)}{1/N^2}}
\end{equation}
with $\partial_m V(m;\alpha,H)=j(m;\alpha,H)$. For $H=0$, the flux $j$ is an odd function in $m$, \Eref{j_useful} and comment after \Eref{nontrivial_roots}, with $j'(0;\alpha,0)$ changing sign at $\alpha=\alpha_c$, so that the potential may be assumed to be of the form
\begin{equation}
    V(m;\alpha,0) = \half r (\alpha_c-\alpha) m^2 + \frac{u}{4!} m^4
\end{equation}
which is enough to extract the form of the variance of $m$ at $\alpha=\alpha_c$ from
\begin{equation}
    \int \dint{m} m^n \Exp{-\frac{(2/N)V(m;\alpha_c,H)}{1/N^2}} \propto 
N^{-\frac{n+1}{4}} \ .
\end{equation}
Together with the normalisation of $\PC_0(m;\alpha,0)$ the variance of $m$ thus scales like $N^{-1/2}$, so that the expectation of $|m|$ will scale like $N^{-1/4}$, namely like mean-field theory, predicting the scaling $L^{-\beta_{\MFT}/\nu_{\MFT}}$ but at $d=4$, with $\beta_\MFT=1/2$, $\nu_\MFT=1/2$ and $N=L^d$. 
Similarly, we find $\chi\propto N^{1/2}$, as $\chi$ is $N$ times the variance, \Eref{ture_susceptibility}. When the random-neighbour behaviour is displayed on a two-dimensional lattice, $N=L^2$, one thus observes finite size scaling $\ave{|m|}\propto L^{-1/2}$ and $\chi\propto L$.
The scaling behaviour is confirmed by numerical simulations conducted directly of the \emph{random neighbour Ising model}, \Fref{random_Ising_m_and_chi}, and the \emph{off-lattice Brownian $q=2$ Potts model} with a sufficiently large effective diffusion constant $v_0$, \Fref{dim_v2000_m_and_chi}.

\begin{figure}[!htb]
\subfloat[\flabel{FSS_mag_random}Finite size scaling of the magnetization $\ave{|m|}$.]
{\includegraphics[width=0.5\columnwidth]{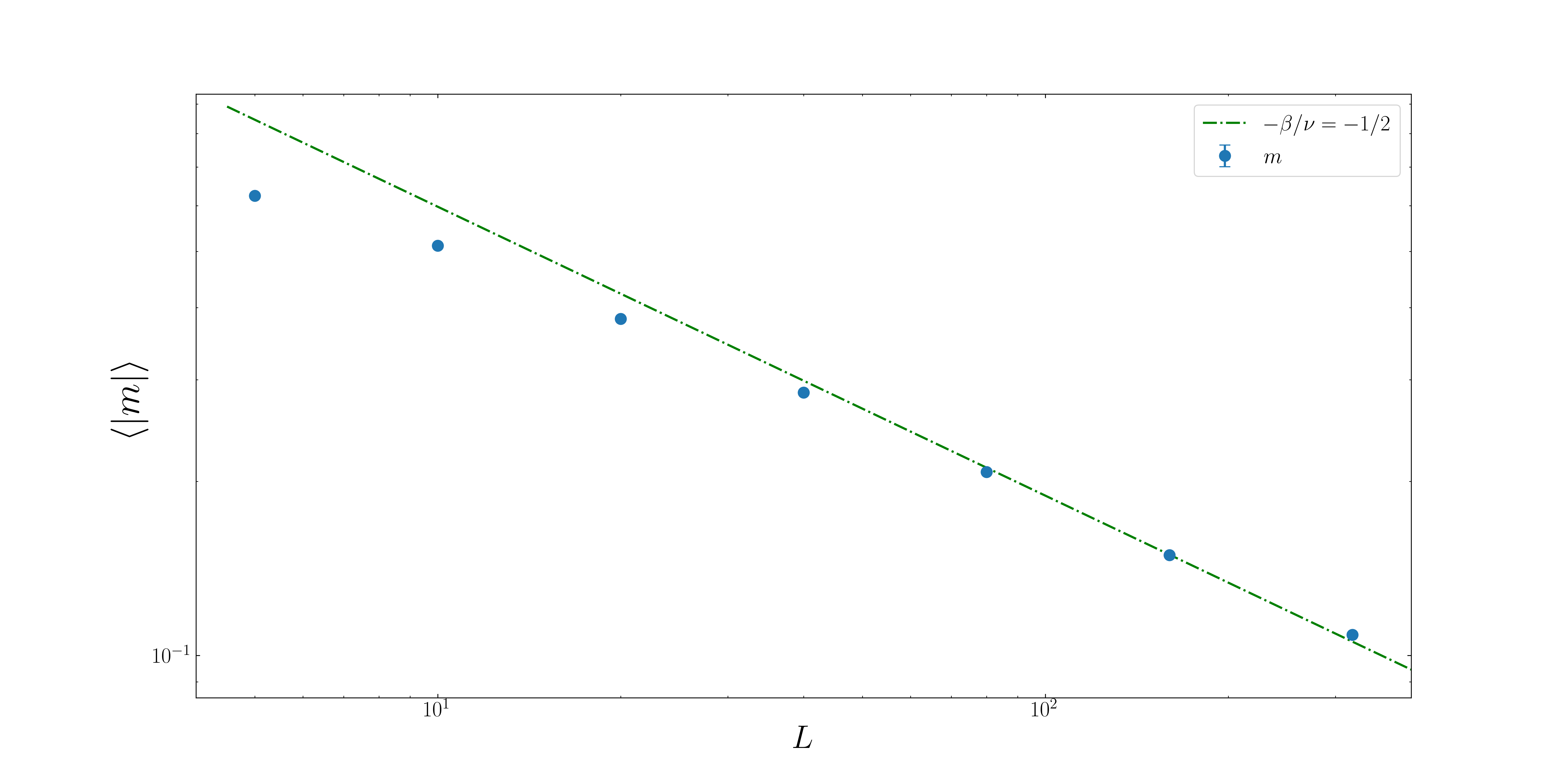}}
\subfloat[\flabel{FSS_chi_random} Finite size scaling of the susceptibility $\chi$.]
{\includegraphics[width=0.5\columnwidth]{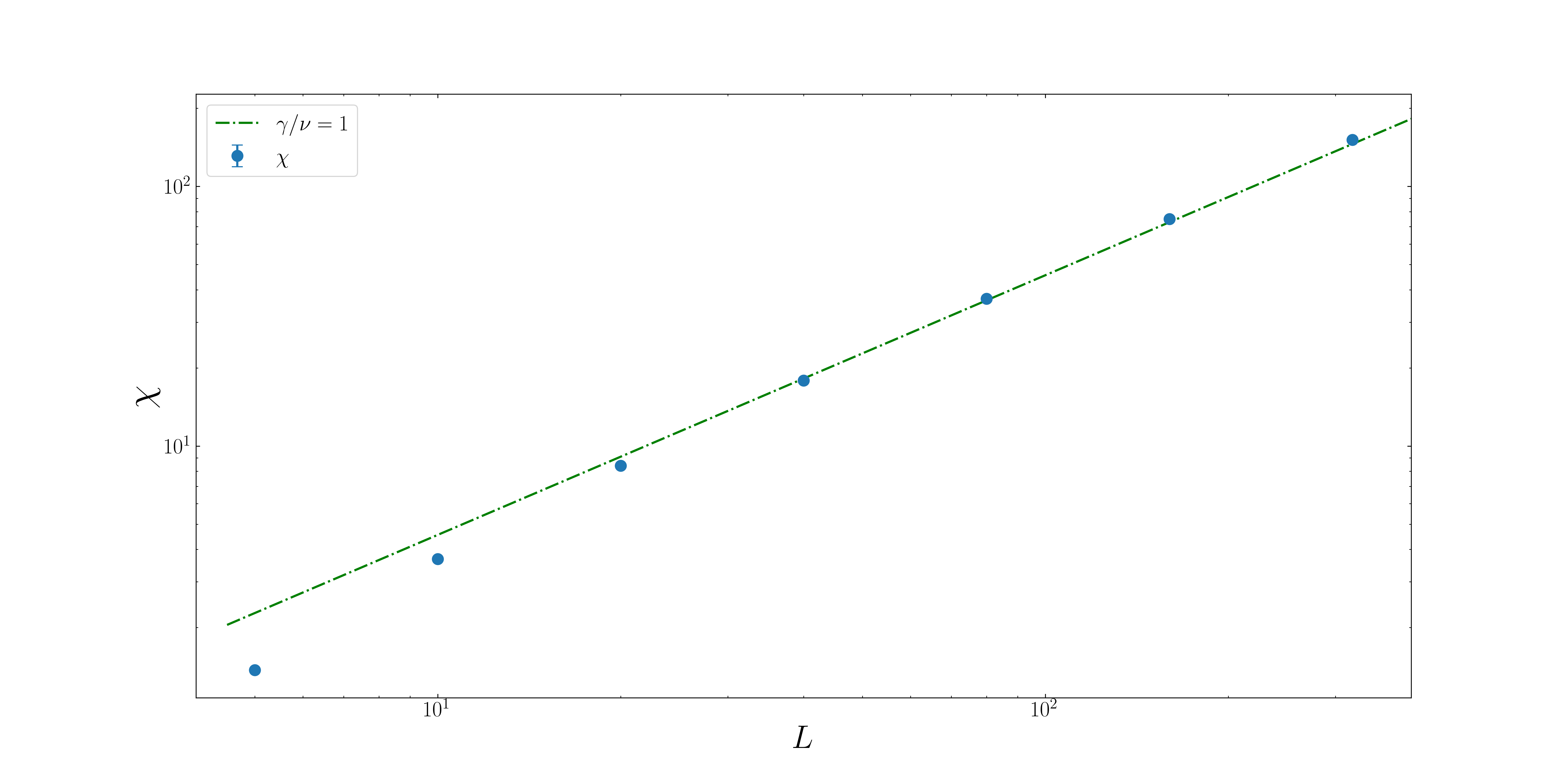}}
\caption{\flabel{random_Ising_m_and_chi} Finite size scaling of the magnetization  $\ave{|m|}$ and the susceptibility $\chi$ at $\invTempCrit$ for random neighbour Ising model with system sizes $L$ ranging from $10$ to $320$. The effective critical temperature was determined by the maximum of the susceptibility, \Sref{observables}.}
\end{figure}

\begin{figure}[!htb]
\subfloat[\flabel{FSS_mag_v2000} Finite size scaling of the magnetization  $\ave{|m|}$.]
{\includegraphics[width=0.5\columnwidth]{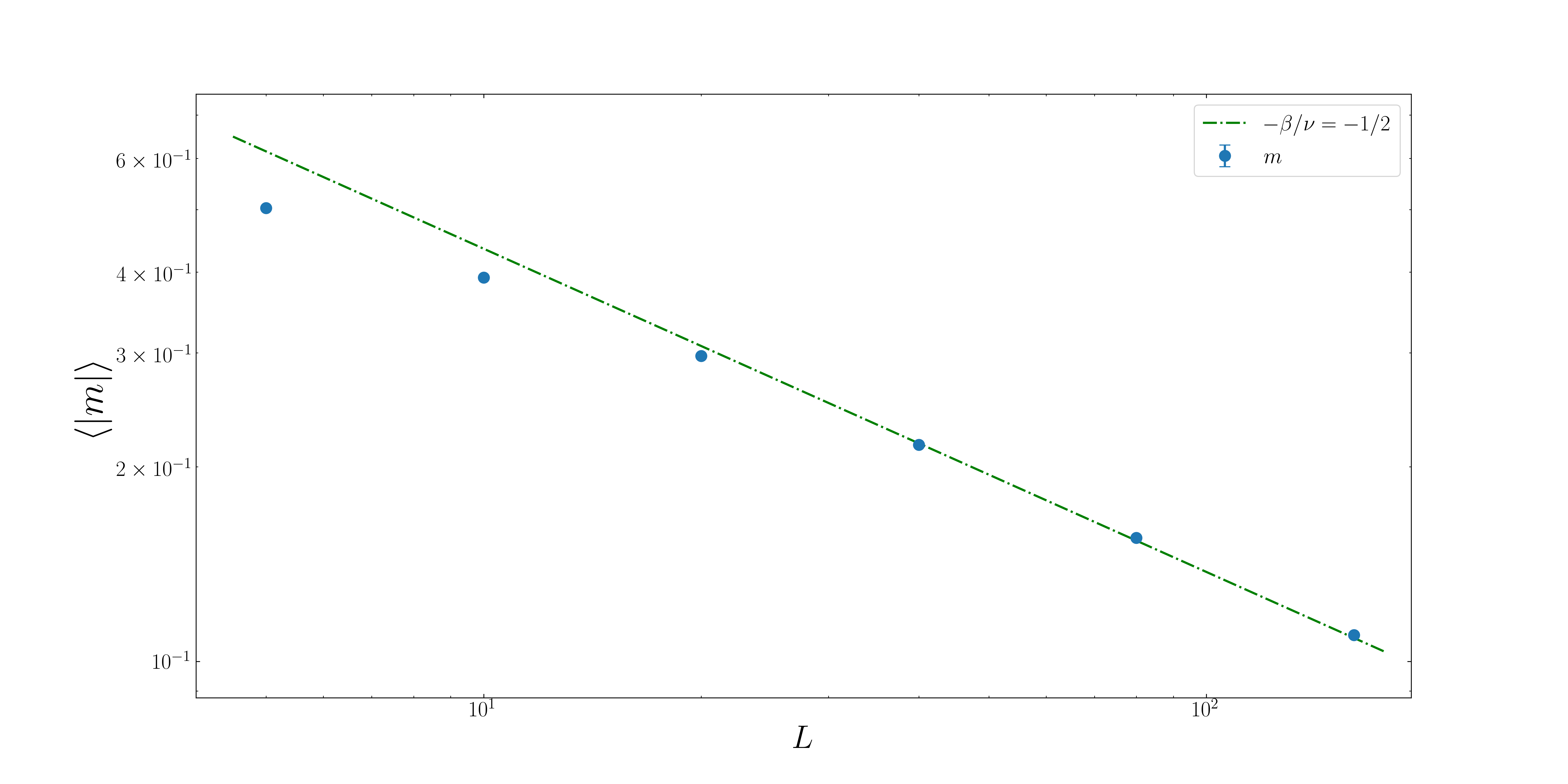}}
\subfloat[\flabel{FSS_chi_v2000} Finite size scaling of the susceptibility $\chi$.]
{\includegraphics[width=0.5\columnwidth]{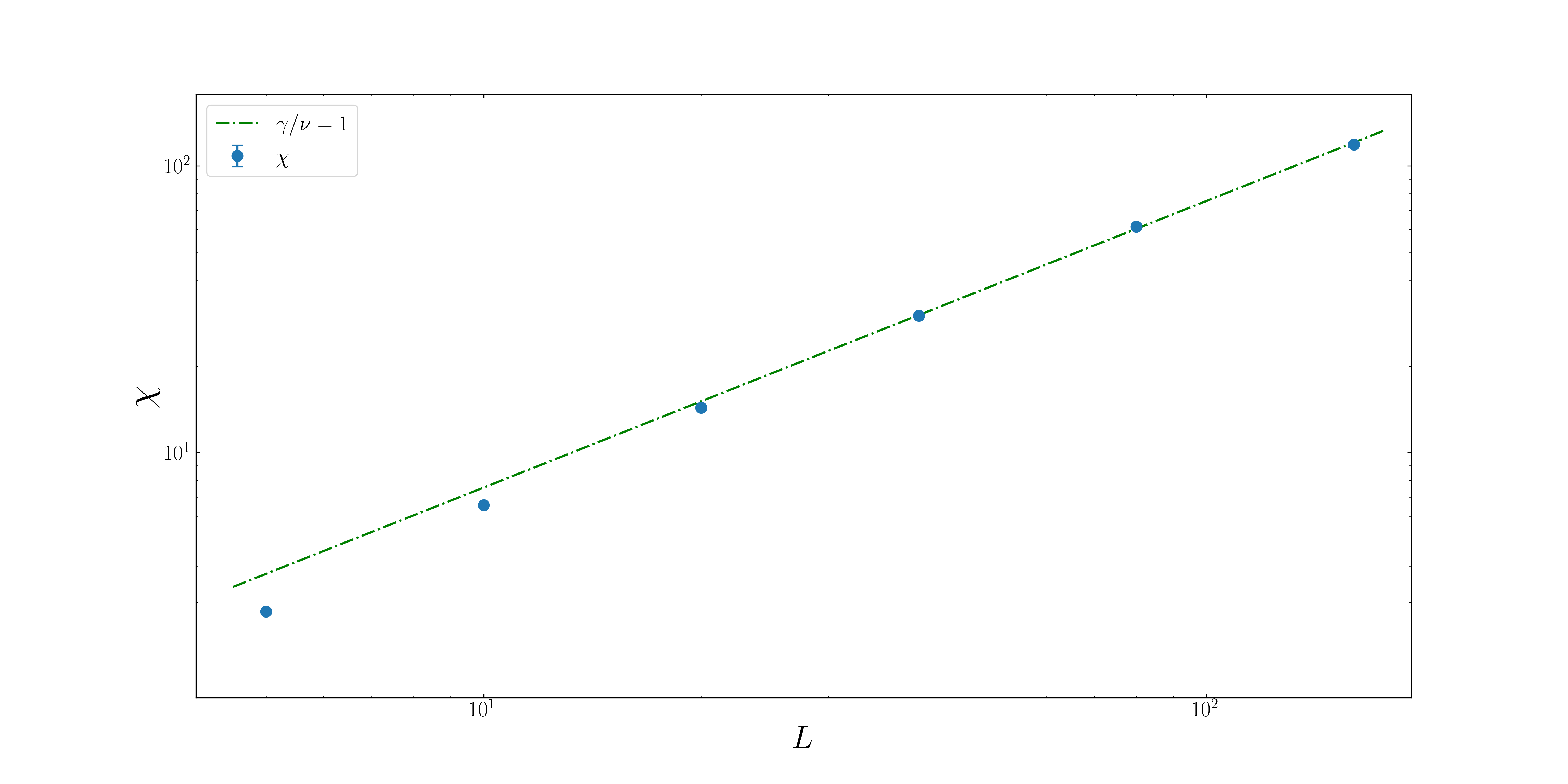}}
\caption{\flabel{dim_v2000_m_and_chi} Finite size scaling of the magnetization  $\ave{|m|}$ and the susceptibility $\chi$ at $\invTempCrit$ for off-lattice Brownian $q=2$ Potts model with system sizes $L$ ranging from $10$ to $160$, with particle density $\rho =2$, interaction range $r_0=1$ and hopping length $v_0=2000$.}
\end{figure}

\subsection{Concrete expressions for $W(m;\alpha,H$)}\seclabel{concrete_W}
To support the results above, we determine the concrete form of $W(m;\alpha,H)$, \Eref{def_W}, for some coordination numbers $q$. The key ingredient is $C_{i}(N^+)$, the probability to choose $i$ up spins when $N^+$ of $N$ spins are in the up state. If the drawing is done with replacement, 
\begin{equation}\elabel{C_with_replacement}
    C_{i}(N^+)=
    \left(\frac{N^+}{N}\right)^i 
    \left(\frac{N-N^+}{N}\right)^{q-i}
    \binom{q}{i}
\end{equation}
and without replacement
\begin{equation}\elabel{C_without_replacement}
    C_{i}(N^+)=
    \frac{(N^+)_i (N-N^+)_{q-i}}{(N)_q}
    \binom{q}{i}
\end{equation}
where $(N)_q$ denotes the falling factorial consisting of $q$ terms, say $(N)_3=N(N-1)(N-2)$.
Both of the probabilities \Erefs{C_with_replacement} and \eref{C_without_replacement} sum to unity, $\sum_{i=0}^q C_{i}(N^+)=1$. 

By inspection of \Eref{C_without_replacement}, replacement  amounts solely to a correction of order $N^{-1}$ compared to \Eref{C_with_replacement}, which we use in the following. Evaluating \Eref{def_W} with the help of \Erefs{def_pplus} and \eref{C_with_replacement} then gives 
\begin{equation}\elabel{W_in_m}
W(m;\alpha,H) = \left(\frac{1}{2}\right)^q 
\sum_{i=0}^q \binom{q}{i} (1+m)^i (1-m)^{q-i}
    \frac
    {\exp{-\alpha(2i-q)+2\beta H}}
    {2\cosh(\alpha(2i-q)+2\beta H)}
\end{equation}
which may be rewritten as
\begin{equation}\elabel{cleaner_W}
    W(m;\alpha,H) = \left(\frac{1}{2}\right)^q (1-m^2)^{q/2}
    \sum_{i=0}^q \binom{q}{i}
    \frac
    {\exp{(\sigma-\alpha)(2i-q)+2\beta H}}
    {2\cosh(\alpha(2i-q)+2\beta H)}
\end{equation}
where 
\begin{equation}
    \exp{\sigma} = 
    \sqrt{
    \frac{1+m}{1-m}
    }
\end{equation}
has been chosen to conveniently absorb powers of $1\pm m$ into the exponential. The precise form of $W(m;\alpha,H)$ strongly depends on the number $q$ of random neighbours.

Using the symmetry of $W$ in \Eref{cleaner_W} under $i\to q-i$ at $H=0$, it can be further simplified,

\begin{subequations}
\elabel{W_general_q}
\begin{align}
W(m;\alpha,0) & =
\left(\frac{1}{2}\right)^q (1-m^2)^{q/2}
\left(
\half \binom{q}{q/2}
+ 
\sum_{k=1}^{q/2}
\binom{q}{k+q/2}
\frac
{\cosh(2k(\alpha-\sigma))}
{\cosh(2k\alpha)}
\right)
&& \text{for $q$ even}\\
W(m;\alpha,0) & =
\left(\frac{1}{2}\right)^q (1-m^2)^{q/2}
\sum_{k=1}^{(q+1)/2}
\binom{q}{k+(q-1)/2}
\frac
{\cosh((2k-1)(\alpha-\sigma))}
{\cosh((2k-1)\alpha)}
&& \text{for $q$ odd.}
\end{align}
\end{subequations}
We may now consider different values of the coordination number $q$.
Formally, $q=0$: In this case $W=\half$ and the flux, \Eref{j_useful}, is $j(m; \alpha, H)=m/2$, causing increase or decrease of $m$ by randomly selecting an up-spin or a down-spin and choosing a new orientation with equal probabilities. $q=1$: $W(m;\alpha,0)=(1-m \tanh(\alpha))/2$ is linear in $m$, so that \Eref{nontrivial_roots} is only ever satisfied for $m=0$. 
$q=2$: $W(m;\alpha,0)=(1-m \tanh(2\alpha))/2$, differing from $q=1$ only by the argument of the $\tanh$.
$q=3$: $W(m;\alpha,0)=\big(1-(3m/4)(1-m^2)\tanh(\alpha)-(m/4)(3+m^2)\tanh(3\alpha)\big)/2$.
For any $q>2$ the current is non-linear in $m$, therefore in principle allowing for non-trivial roots \Eref{nontrivial_roots}. 

We can use the explicit expressions for $W(m;\alpha,0)$ to confirm the considerations around \Eref{m_solns} finding for $q=3$ that $W^{(3)}(0;\alpha,0)>0$ for all $\alpha>0$, that $\partial_m|_{m=0}j(m;\alpha,0)$ becomes negative when $\partial_m|_{m=\mbar}j(m;\alpha,0)$ positive and that $\mbar$ as of \Eref{m_solns} grows initially like $\sqrt{\alpha}$.

\subsection{Off-lattice}\seclabel{Off_lattice_random}
One may extend the results above by considering particles which have a random number of neighbours. This would be the case, if the particles are quickly diffusing, so that a random number of them need to be considered to be within a supposed interaction range. This process can be approximated by a spatial Poisson process with density $\rho$ in an area $A$, so that the probability of finding $q$ interaction partners is $\pi_q=\exp{-\rho A}(\rho A)^q/q!$. That this is an approximation can be seen as $\pi_q>0$ even for $q\ge N$.

The effective flux of the present setup may be written as
\begin{equation}
j_\eff(m; \alpha, H)=
\half m - \half + \sum_{q=0}^\infty \pi_q W_q(m; \alpha, H)
\end{equation}
using \Eref{j_useful} and evaluating the right-hand side using the general form of $W(m; \alpha, H)$, \Eref{W_in_m}, now sub-scripted by $q$ to emphasise the dependence on $q$. As seen above, for any $q>2$, the flux $j$ is non-linear in $m$, in fact $W_q(m;\alpha,0)=\half+\OC(m^q)$, where $\OC(m^q)$ (in small $m$) has to be odd in $m$, so that $j_\eff(m;\alpha,0)=\half m + (\text{polynomial odd in $m$})$. The polynomial contains coefficients that remain bounded for any $\alpha$, so that small values of $\rho$ entering into $\pi_q$ put so much weight on $q=0,1,2$, that $j_\eff$ has only a trivial root. In other words: At sufficiently low particle density, the off-lattice random neighbour model Ising Model cannot display order at any temperature.

The connection to the Potts Model is somewhat subtle, because the variation of the number of interaction partners, but as commented above, \Sref{off_lattice_particle}, these fluctuations do not enter in any of the dynamics, so that Ising and Potts Model as discussed here show the same behaviour at $2T$ and $T$ respectively.

\end{widetext}
\end{document}